\PassOptionsToPackage{table,xcdraw}{xcolor}
\documentclass[12pt,a4paper]{article}
\pdfoutput=1

\usepackage{geometry}
\usepackage[numbers,sort&compress]{natbib}
\usepackage{amsmath}
\usepackage{amssymb}
\usepackage[dvips]{graphicx}
\usepackage{pstricks}
\usepackage{bm}
\usepackage{pbox}
\usepackage{placeins}
\usepackage{graphicx}
\usepackage{caption}
\usepackage{subcaption}
\usepackage[T1]{fontenc}
\usepackage{footnote}
\usepackage{pdfpages}
\usepackage{hhline}
\usepackage{multirow}
\usepackage{enumitem}
\usepackage{slashed}
\usepackage[nottoc,notlot,notlof]{tocbibind}
\usepackage[title,titletoc]{appendix}
\usepackage{dsfont}
\usepackage{pifont}
\allowdisplaybreaks
\usepackage{cases}
\usepackage{tikz}
\usepackage[symbol]{footmisc}

\renewcommand{\thetable}{\Roman{table}}
\newcommand\scalemath[2]{\scalebox{#1}{\mbox{\ensuremath{\displaystyle #2}}}}

\definecolor{MyDarkBlue}{rgb}{0.1, 0.1, 0.8}
\definecolor{MyLightBlue}{rgb}{0.22,0.51,0.9}
\definecolor{MyGreen}{rgb}{0.0, 0.5, 0.0}
\definecolor{BrickRed}{rgb}{0.8, 0.25, 0.33}
\usepackage[colorlinks=true,linkcolor=blue,citecolor=MyDarkBlue,
urlcolor=MyLightBlue,bookmarksnumbered=true,bookmarksopen]{hyperref}
\hypersetup{colorlinks, citecolor=blue,linkcolor=blue, urlcolor=MyLightBlue}

\begin{document}
\vspace*{-0.2in}
\begin{flushright}

\end{flushright}
\vspace{0.5cm}
\begin{center}
{\Large\bf
Probing Minimal Grand Unification through Gravitational Waves, Proton Decay, and Fermion Masses
}
\end{center}

\vspace{0.5cm}
\renewcommand{\thefootnote}{\fnsymbol{footnote}}
\begin{center}
{\large
{}~\textbf{Shaikh Saad}\footnote[2]{E-mail:  
\textcolor{MyLightBlue}{shaikh.saad@unibas.ch}
}
}
\vspace{0.5cm}

{\em Department of Physics, University of Basel, Klingelbergstrasse\ 82, CH-4056 Basel, Switzerland}
\end{center}

\renewcommand{\thefootnote}{\arabic{footnote}}
\setcounter{footnote}{0}
\thispagestyle{empty}
\begin{abstract}
Motivated by the direct discovery of gravitational waves (GWs) from black holes and neutron stars, there is a growing interest in investigating GWs from other sources. Among them, GWs from cosmic strings are particularly fascinating since they naturally appear in a large class of grand unified theories (GUTs). Remarkably, a series of pulsar-timing arrays (PTAs) might have already observed GWs in the nHz regime, hinting towards forming a cosmic string network in the early universe, which could originate from phase transition associated with the seesaw scale emerging from GUT. In this work, we show that if these observations from PTAs are confirmed, GWs from cosmic strings, when combined with fermion masses, gauge coupling unification, and proton decay constraints, the parameter space of the minimal SO(10) GUT becomes exceedingly restrictive. The proposed minimal model is highly predictive and will be fully tested in a number of upcoming gravitational wave observatories.  
\end{abstract}
\newpage
{\hypersetup{linkcolor=black}
\tableofcontents}
\setcounter{footnote}{0}

\section{Introduction}
The basic idea of grand unified theories (GUTs)~\cite{Pati:1973rp,Pati:1974yy,Georgi:1974sy, Georgi:1974yf, Georgi:1974my, Fritzsch:1974nn,Langacker:1980js} is embedding the Standard Model (SM) gauge group in a simple non-Abelian group.  Consequently, the three gauge couplings associated with the SM gauge group $SU(3)_C\times SU(2)_L\times U(1)_Y$ that have different values at the low energies are unified into a single gauge coupling at a very high energy scale, namely, the GUT scale, $M_\textrm{GUT}$.  In this setup, all the SM fermions of each generation can also be unified within a single representation under the chosen GUT group. 

The minimal simple group that can serve the aforementioned purposes is the $SO(10)$ group~\cite{Georgi:1974my, Fritzsch:1974nn}. Interestingly, the $16$-dimensional spinorial representation of  $SO(10)$ group, in addition to  all the SM fermions, contains a right-handed neutrino $\nu_R$, which via seesaw mechanism can provide the SM neutrinos a tiny non-zero masses.  To evade the current proton decay constraints, the GUT breaking scale has to be quite large, roughly, $M_\textrm{GUT}\gtrsim 10^{16}$ GeV, whereas explaining the light neutrino oscillation data requires the right-handed neutrinos to have masses much smaller than the GUT scale that fixes the  intermediate  breaking scale to be of order  $v_R\sim 10^{12}-10^{14}$ GeV.

Remarkably, this intermediate stage symmetry breaking that generates the right-handed neutrino masses   by breaking the $B-L$ ($U(1)_{B-L}$) by two units leads to the formation of a cosmic string network in the early Universe~\cite{Vilenkin:1984ib,Caldwell:1991jj,Hindmarsh:1994re}. Among all the topological defects formed during the subsequent breaking of the GUT symmetry to the SM group, cosmic strings are  particularly interesting because, unlike monopoles and domain walls that are ruled out, the existence of cosmic strings has compelling observational consequences.  These cosmic strings intercommunicate and eventually emit gravitational radiation forming a
stochastic gravitational wave (GW) background that ongoing experiments  and future GW interferometers can test.

Inspired by the direct discovery of
GWs from black holes and neutron stars by the LIGO and Virgo Collaborations~\cite{LIGOScientific:2016aoc,LIGOScientific:2016sjg,LIGOScientific:2017bnn,LIGOScientific:2017vox,LIGOScientific:2017ycc,LIGOScientific:2017vwq,LIGOScientific:2020zkf,LIGOScientific:2020iuh} at frequencies $f\gtrsim 10$ Hz, there is widespread interest in experiments exploring GW signals in other frequency ranges. Among others, Pulsar-timing arrays (PTAs) are sensitive to GW signals with frequencies in the nHz range and ideal for detecting GWs originating from cosmic strings. Intriguingly, pulsar-timing arrays:  NANOGrav~\cite{NANOGrav:2020bcs}, PPTA~\cite{Goncharov:2021oub}, EPTA~\cite{Chen:2021rqp}, and IPTA~\cite{Antoniadis:2022pcn}  might
have already seen evidence of a primordial gravitational
wave background that could originate from the breaking of  the seesaw scale    emerging from $SO(10)$ GUT. Moreover,  determining the intermediate symmetry breaking scale from neutrino oscillation data as well as  from gravitational
wave signal strongly constrains  the GUT scale, which subsequently has a profound effect on determining the proton lifetime.

This work considers a minimal grand unified theory based on $SO(10)$ gauge symmetry. The presence of no other global or discrete symmetries is assumed. Within this framework, the most minimal Yukawa sector consists of a real $10_H$, a real $120_H$, and a complex $126_H$ dimensional representations~\cite{Babu:2016bmy}. The Yukawa sector of this theory has only a limited number of parameters to reproduce the charged fermion and neutrino masses and mixing angles. Owing to the minimality criterion, we restrict ourselves to small representations and employ a $45_H$ and a $54_H$ dimensional Higgs representations to break the GUT symmetry. By acquiring vacuum expectation values (VEVs), the SM singlet components of  $45_H+54_H$ fields break the GUT symmetry to an intermediate symmetry that contains $U(1)_{B-L}$ symmetry. Subsequent breaking by $126_H$ multiplet gives super-heavy masses to the right-handed neutrinos and is responsible for the formation of the cosmic string network. Our detailed study reveals that only a small range for the seesaw scale is permitted when constraints from GWs detectors, SM fermion mass spectrum, gauge coupling unification, and proton decay are imposed, making the proposed model highly predictive and fully testable.

This paper is organized in the following way: in Secs.~\ref{sec:yukawa} and ~\ref{sec:fit} we discuss the details of the Yukawa sector and perform numerical fits. In Sec.~\ref{sec:breaking}, symmetry breaking chain and the associated topological defects are elaborated. Gravitational waves from cosmic string networks is discussed in Sec.~\ref{sec:string}. Furthermore, unification and proton decay constraints are examined in Sec.~\ref{sec:PDunification}. Finally, we conclude in Sec.~\ref{sec:conclusions}.

\section{The Yukawa sector}\label{sec:yukawa}
In a renormalizable $SO(10)$ theory with no new fermions beyond the three families of chiral 16s,  the Higgs multiplets that can contribute to the fermion masses can be identified from the following fermion bilinears:
\begin{equation}
16\times 16 = 10_s+ 120_a + 126_s.
\end{equation}
The 10 and the 120 are real representations in $SO(10)$, whereas the 126 is complex. In the above equation, the subscripts $s$ and $a$ stand for symmetric and antisymmetric components in the family space. With this, the most general renormalizable Yukawa sector takes the following form: 
\begin{equation}
\label{yukawa}
\mathcal{L}_{yuk}= 16_F (Y_{10}10_H+Y_{120}120_H+Y_{126}\overline{126}_H) 16_F.
\end{equation}
It turns out that the Yukawa sector of  $SO(10)$ GUTs are very predictive, which have been extensively analyzed in the literature~\cite{Babu:1992ia, Bajc:2001fe,Bajc:2002iw,Fukuyama:2002ch,Goh:2003sy,
Goh:2003hf,Bertolini:2004eq, Bertolini:2005qb, Babu:2005ia,Bertolini:2006pe, Bajc:2008dc,
Joshipura:2011nn,Altarelli:2013aqa,Dueck:2013gca, Fukuyama:2015kra, Babu:2016cri, Babu:2016bmy, Saad:2017wgd, Babu:2018tfi, Babu:2018qca, Ohlsson:2019sja,Babu:2020tnf,Mummidi:2021anm}.

The up-quark, down-quark, charged leptons, Dirac neutrino, and Majorana neutrino mass matrices derived from Eq.~\eqref{yukawa} can now be written down~\cite{Babu:2016bmy}:
\begin{align}
&M_U= \underbrace{v_{10}Y_{10}}_{\equiv D}+\underbrace{v^u_{126}Y_{126}}_{\equiv S}+\underbrace{(v^{(1)}_{120}+v^{(15)}_{120})Y_{120}}_{\equiv A},\label{Mu}\\
&M_D= v^{\ast}_{10}Y_{10}+v^d_{126}Y_{126}+(v^{(1)\ast}_{120}+v^{(15)\ast}_{120})Y_{120},\\
&M_E= v^{\ast}_{10}Y_{10}-3v^d_{126}Y_{126}+(v^{(1)\ast}_{120}-3v^{(15)\ast}_{120})Y_{120},\\
&M_{\nu_D}= v_{10}Y_{10}-3v^u_{126}Y_{126}+(v^{(1)}_{120}-3v^{(15)}_{120})Y_{120},\\
&M_{\nu_{R}}=v_{R}Y_{126}. \label{MR}
\end{align}
One crucial feature of the above set of mass matrices is that the same factor of $|v_{10}|$ appears both in the up (up-type quark and Dirac neutrino) and the down (down-type quark and charged lepton) sectors. Since $10_H$ is a real field in this model, the Higgs doublet residing inside $10_H$ is self-conjugate. As a result, the bi-doublet $\Phi=(1,2,2)\subset 10_H$ under the Pati-Salam group satisfies the following condition,
\begin{align}
&\Phi^*=\tau_2\Phi\tau_2    
\\
\mathrm{i.e.,}\;&\Phi=\begin{pmatrix}
\phi_0&\phi^+\\
\phi^-&\phi_0^*
\end{pmatrix},
\end{align}
leading to identical vacuum expectation values both in the up and the down sectors, $v_u=v^*_d=v_{10}$.

By defining
\begin{align}
r_1&=\frac{v^d_{126}•}{•v^u_{126}},\;\; r_2=\frac{•v^{(1)\ast}_{120}-3v^{(15)\ast}_{120}}{•v^{(1)}_{120}+v^{(15)}_{120}},\;\; e^{i\phi}= \frac{•v^{(1)\ast}_{120}+v^{(15)\ast}_{120}}{•v^{(1)}_{120}+v^{(15)}_{120}},\;\; c_{R}=\frac{•v_{R}}{•v^u_{126}},\label{cR}
\end{align}
we re-write the above mass matrices as,
\begin{align}
&M_{U}= D+S+A\equiv vY_U, \label{E1}\\
&M_{D}= D+r_1 S+ e^{i \phi} A\equiv vY_D, \label{MD}\\
&M_{E}= D-3 r_1 S+ r_2 A\equiv vY_E, \label{ME} \\
&M_{\nu_D}= D-3 S+ r^{\ast}_2 e^{i \phi} A\equiv vY_{\nu_D}, \label{MD}\\
&M_{\nu_{R}} = c_{R} S . \label{E2}
\end{align}
The mass matrices are written in a basis $f^c_iM_{ij}f_j$, and we have chosen a phase convention
where $v_{10}$ is made real by an $SU(2)_L$ rotation. Without loss of generality, we choose to work in a basis where the matrix $S$ is diagonal. As usual, $v=174.104$ GeV.  When charged fermion Yukawa coupling matrices are bi-diagonalized, we denote the corresponding diagonal couplings as  $Y^\textrm{diag}_U=\left(y_u,y_c,y_t\right)$, $Y^\textrm{diag}_D=\left(y_d,y_s,y_b\right)$, and $Y^\textrm{diag}_E=\left(y_e,y_\mu,y_\tau\right)$, for the up-type, down-type quarks, and charged lepton sectors, respectively.   The light neutrino mass matrix, obtained from the seesaw formula~\cite{Minkowski:1977sc,Yanagida:1979as,Glashow:1979nm,GellMann:1980vs,Mohapatra:1979ia}, assuming the typical type-I dominance, is given by
\vspace{-5pt}
\begin{align}
\label{MN}
M_N = - M^T_{\nu_D} M^{-1}_{\nu_R} M_{\nu_D} .
\end{align}
Due to minimal number of parameters, in this work, our focus is on type-I dominance.  A full exploration of the type-I+II case is left for a future exploration.

It is crucial to identify the intermediate symmetry breaking scale for our analysis.  First, note that the mass of the heaviest right-handed neutrino is given by $M_3=v_R (Y_{126})_{33}$, see Eq.~\eqref{MR}. Moreover, this model has the specific property that the top quark mass is solely determined by the matrix $S$  that provides $m_t=v^u_{126} (Y_{126})_{33}$, see Eq.~\eqref{Mu}.   This can be explained as follows. First, note that at the GUT scale $m_t/m_b\sim 75$ (see Table~\ref{result}). Moreover, the dominant contribution to the bottom-quark mass in Eq.~\eqref{MD} comes from $\left(D_{33}+r_1S_{33}\right)$, which demands  $\left(D_{33}+r_1S_{33}\right)/S_{33}\sim 10^{-2}$ to correctly reproduce the above-mentioned ratio between the top and the bottom quark masses. Hence, $D_{33}, r_1S_{33}\ll S_{33}\sim m_t$, as anticipated.     As will be shown later, the GUT scale value of the top quark mass is $m_t\approx 85$ GeV. Consequently, the lowest value of $(Y_{126})_{33}$ corresponds to the maximum value allowed for $v^u_{126}\lesssim 174$ GeV, which provides $(Y_{126})_{33}^\mathrm{low}=0.488$. On the other hand, to ensure perturbativity of the Yukawa couplings, we allow the highest value of  $(Y_{126})_{33}^\mathrm{high}=2$.  Since the mass of the heaviest right-handed neutrino is given by $M_3=v_R (Y_{126})_{33}$, the lowest and the highest symmetry breaking scales producing the cosmic string network  are determined to be
\begin{align}
v_R=\begin{cases}
    v_R^\textrm{min}= 0.5\times M_3.\\
    v_R^\textrm{max}= 2.05\times M_3.
  \end{cases}\label{vR:range}    
\end{align}
As will be discussed, a fit to fermion mass spectrum prefers a large $M_3$ value, hence is highly constrained from GW detectors.  (A choice of $(Y_{126})_{33}^\mathrm{high}=\sqrt{4\pi}$ would lead to $v_R^\textrm{min}= 0.28\times M_3$.)

\section{Fit to the fermion masses and mixings}\label{sec:fit}
In this section, we provide numerical fits to fermion masses and mixing utilizing the mass matrices derived in the previous section. We substantially improve the fitting procedure compared to Ref.~\cite{Babu:2016bmy}.  Our fitting procedure is summarized as follows: 
\begin{enumerate}
\item The fermion mass matrices Eqs.~\eqref{E1}-\eqref{MN} contain only a limited number of parameters: 15 magnitudes and 12 phases.  All these parameters are randomly varied at the GUT scale, which for fitting purposes, we fix at $M_{GUT}=2\times 10^{16}$ GeV.

\item A significant improvement over the previous study is that we take into account the full SM+Type-I seesaw renormalization group equation (RGE) running from $M_{GUT}$ to $M_Z$ scale. For this task, we utilize the package \texttt{REAP}~\cite{Antusch:2005gp}, which integrates out  heavy right-handed neutrinos   successively at their respective mass
thresholds. This is very crucial due to the extremely hierarchical mass structure in the model under consideration. Since the intermediate scale is quite close to the GUT scale, to the precision we are working, it is good enough to consider only  SM+Type-I RGEs. Incorporating full-fledged RGEs from the intermediate to the GUT scale, which  is expected to provide a small correction, is, however, beyond the scope of this work.

\FloatBarrier
\begin{table}[th!]
\centering
\footnotesize
\resizebox{0.99\textwidth}{!}{
\begin{tabular}{|c||c|c|c||c|c|}
\hline
\textbf{Observables} & \multicolumn{3}{c|}{Values at $M_Z$ scale} & \multicolumn{2}{c|}{Values at \textbf{$M_{GUT}$} scale}   \\ 

\cline{2-6}
($\Delta m^2_{ij}$ in $eV^2$) &Input&Best Fit: NO& Best Fit: IO &   
 NO&IO 
\\
\hline \hline

\rowcolor{red!10}$y_u/10^{-6}$&6.65$\pm$2.25&6.65&6.71&  2.85&2.87\\  
\rowcolor{red!10}$y_c/10^{-3}$&3.60$\pm$0.11&3.60&3.60&  1.54&1.54\\ 
\rowcolor{red!10}$y_t$&0.986$\pm$0.0086&0.986& 0.986&  0.48&0.48\\ 

\rowcolor{yellow!10}$y_d/10^{-5}$&1.645$\pm$0.165&1.645&1.675&  0.73&0.74\\ 
\rowcolor{yellow!10}$y_s/10^{-4}$&3.125$\pm$0.165&3.125&3.146&  1.38&1.39\\ 
\rowcolor{yellow!10}$y_b/10^{-2}$&1.639$\pm$0.015&1.639&1.639&  0.637&0.637\\ 

\rowcolor{cyan!10}$y_e/10^{-6}$&2.7947$\pm$0.02794&2.7947&2.7899&  2.8873&2.8817\\ 
\rowcolor{cyan!10}$y_\mu/10^{-4}$&5.8998$\pm$0.05899&5.8998&5.9021&  5.924&5.894\\ 
\rowcolor{cyan!10}$y_\tau/10^{-2}$&1.0029$\pm$0.01002&1.0029&1.0012&  0.985&0.989\\

\rowcolor{orange!10}$\theta_{12}^\textrm{CKM}/10^{-2}$&$22.735\pm$0.072&22.735&22.739&  22.73&22.74\\ 
\rowcolor{orange!10}$\theta_{23}^\textrm{CKM}/10^{-2}$&4.208$\pm$0.064&4.208&4.204&  4.79&4.79\\ 
\rowcolor{orange!10}$\theta_{13}^\textrm{CKM}/10^{-3}$&3.64$\pm$0.13&3.64&3.64&  4.15&4.15\\ 
\rowcolor{orange!10}$\delta^\textrm{CKM}$&1.208$\pm$0.054&1.208&1.204&  1.207&1.204\\

\rowcolor{green!10}$\Delta m^2_{21}/10^{-5} $&7.425$\pm$0.205&7.425&7.433&  9.714&950.84\\ 
\rowcolor{green!10}$\Delta m^2_{31}/10^{-3}$ (NO)&2.515$\pm$0.028&2.515&-&  12.909&-\\
\rowcolor{green!10}$\Delta m^2_{32}/10^{-3}$ (IO)&-2.498$\pm$0.028&-&-2.497&  -&-12.515\\ 

\rowcolor{blue!10}$\sin^2 \theta_{12}$ &0.3045$\pm$0.0125&0.3045&0.3053&  0.308&0.177\\ 
\rowcolor{blue!10}$\sin^2 \theta_{23}$ (NO)$^*$  &0.5705$\pm$0.0205&0.5726&-&  0.484&-\\ 
\rowcolor{blue!10}$\sin^2 \theta_{23}$ (IO)$^*$ &0.576$\pm$0.019&-&0.5819&  -&0.542\\
\rowcolor{blue!10}$\sin^2 \theta_{13}$ (NO)&0.02223$\pm$0.00065&0.02223&-&  0.007&-\\  
\rowcolor{blue!10}$\sin^2 \theta_{13}$ (IO)&0.02239$\pm$0.00063&-&0.02238&  -&0.0223\\

\hline \hline
\rowcolor{gray!10}$\chi^2$&-&$3\times 10^{-8}$&2.77$^\dagger$&-&- \\
\hline

\end{tabular}
}
\caption{Best fit values of the observables without restricting the $U(1)_{B-L}$ breaking scale. ${}^*$Note that experimental measurements of $\theta_{23}$ have two local minimum, one corresponds to $\theta_{23}<45^{\circ}$ and the other $\theta_{23}>45^{\circ}$, see Ref.~\cite{NUFIT}.   The Yukawa couplings, $y_f$, at the GUT scale are defined below Eq.~\eqref{E2}. ${}^\dagger$It is important to point out that for the inverted neutrino mass ordering, a total $\chi^2\approx 2.7$ is entirely coming from $\left(\Delta\chi^2\right)_{\theta_{23}}\approx 2.7$, for details, see Ref.~\cite{NUFIT}.  }\label{result}
\end{table}

\item We perform a minimization of a $\chi^2$-function that includes: 18 observables: 6 quark masses, 3 quark mixing angles,
1 CKM phase, 3 charged lepton masses, 2 neutrino mass squared differences, and 3 mixing
angles in the neutrino sector. Since the Dirac CP phase in the neutrino sector has not been measured yet, we do not include it in the $\chi^2$-function. All these observables are fitted at the $M_Z=91.8176$ GeV scale after running them down from the GUT scale. Low scale measured values of the observables in the charged and neutral fermion sectors are taken from Refs.~\cite{Antusch:2013jca} and~\cite{NUFIT,Esteban:2020cvm}, respectively. 

\item Finally, for a chosen benchmark scenario (with $M_3\sim 4\times 10^{13}$ GeV), a Markov chain Monte Carlo (MCMC) analysis is performed to explore the parameter space.  
\end{enumerate}

To demonstrate the viability of this theory, first, we perform a fit to fermion masses and mixings without restricting the intermediate $U(1)_{B-L}$ breaking scale. The resulting best fit values are presented in Table~\ref{result}.  It is interesting to point out that the minimal $SO(10)$ Yukawa sector admits normal as well as inverted mass ordering for the neutrinos. Allowing a solution to the inverted neutrino mass ordering is a very distinct feature of the current model, which was not realized in the literature when this model was first introduced in Ref.~\cite{Babu:2016bmy}. A collection of additional predicted quantities in the neutrino section are given  in Table~\ref{tab:predictions}. Moreover,  
best fit parameters (at the GUT scale) are given  in Eqs.~\eqref{fit1:NO}-\eqref{fit2:NO}  and in Eqs.~\eqref{fit1:IO}-\eqref{fit2:IO} for normal ordering (NO) and inverted ordering (IO) cases, respectively.

\begin{table}[t!]
\centering
{\footnotesize
\resizebox{0.99\textwidth}{!}{
    \begin{tabular}{|c||c|c|}
    \hline
       {\bf Quantity}  & \multicolumn{2}{c|}{\bf Best fit prediction}  \\
       \cline{2-3}  &NO&IO \\
\hline \hline

$\left(m_1,m_2,m_3\right)$& $\left(0.00014, 0.0086, 0.0501\right)$   eV &$\left(0.04922, 0.04997, 0.00038\right)$   eV     
 \\[3pt]
\hline

$\left(\sum_i m_i,m_\beta,m_{\beta\beta}\right)$  & $\left(0.0589, 0.0088, 0.0016\right)$ eV       & $\left(0.099, 0.041, 0.033\right)$ eV   
 \\[3pt]
\hline

$\left(\delta,\varphi_1,\varphi_2\right)$ & $\left( 326.4, 109.0, 94.7 \right)^\circ$ & $\left( 209.5, 164.4, 343.4 \right)^\circ$        
 \\[3pt]
\hline

$\left(M_1,M_2,M_3\right)$  & $\left( 2.13\times 10^5, 6.46\times 10^{11}, 2.28\times 10^{14} \right)$ GeV     & $\left( 1.31\times 10^4, 6.42\times 10^{11}, 2.37\times 10^{14} \right)$ GeV       
 \\[3pt]
\hline     

    \end{tabular}
    }
    \caption{ Best fit predictions of several quantities in the neutrino sector. We have defined $U_\mathrm{PMNS}=V_\mathrm{PMNS}\; \mathrm{diag}\{e^{-\varphi_1/2},e^{-\varphi_2/2},1  \}$, where $V_\mathrm{PMNS}$ is the CKM like matrix that contains the Dirac phase.  Here  $m_{\beta}=\sqrt{\sum_{i} |U_{e i}|^{2} m_{i}^2}$ is the effective mass parameter for beta decay and $m_{\beta \beta}= | \sum_{i} U_{e i}^{2} m_{i} |$ is the effective mass parameter for neutrinoless double beta decay. }
    \label{tab:predictions}
    }
\end{table}

\begin{figure}[t!]
\centering
\includegraphics[width=0.48\textwidth]{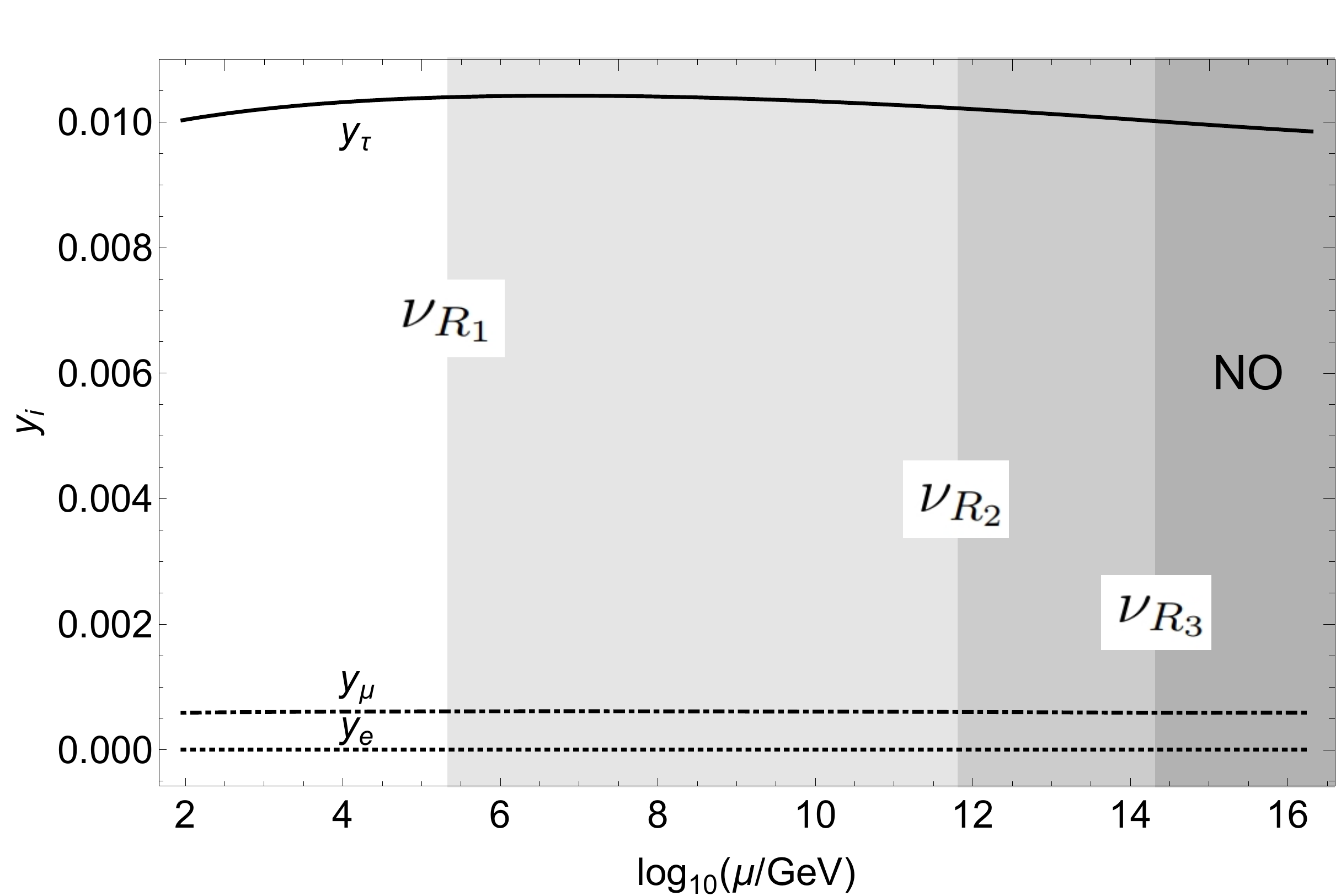}
\includegraphics[width=0.48\textwidth]{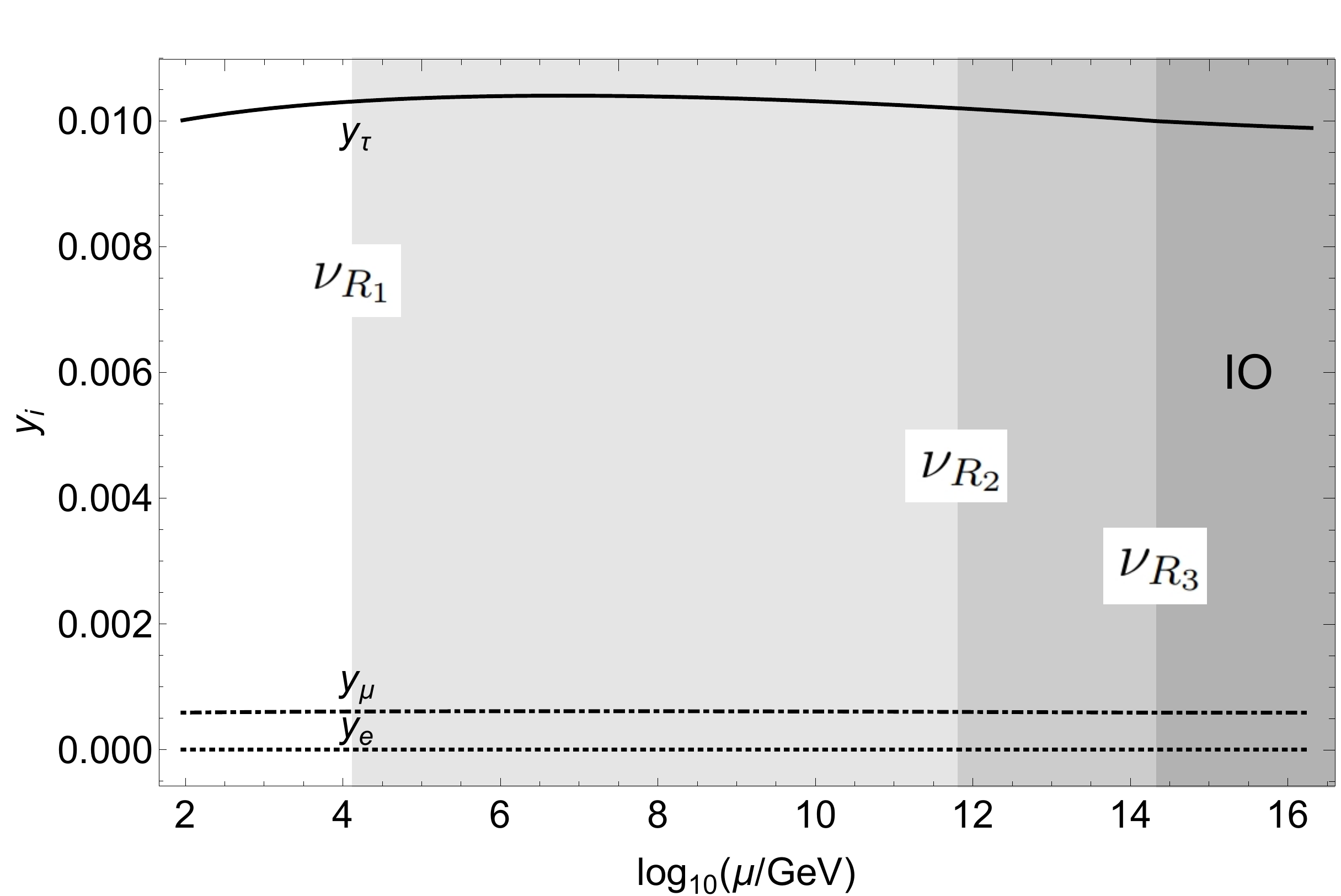}
\\
\includegraphics[width=0.48\textwidth]{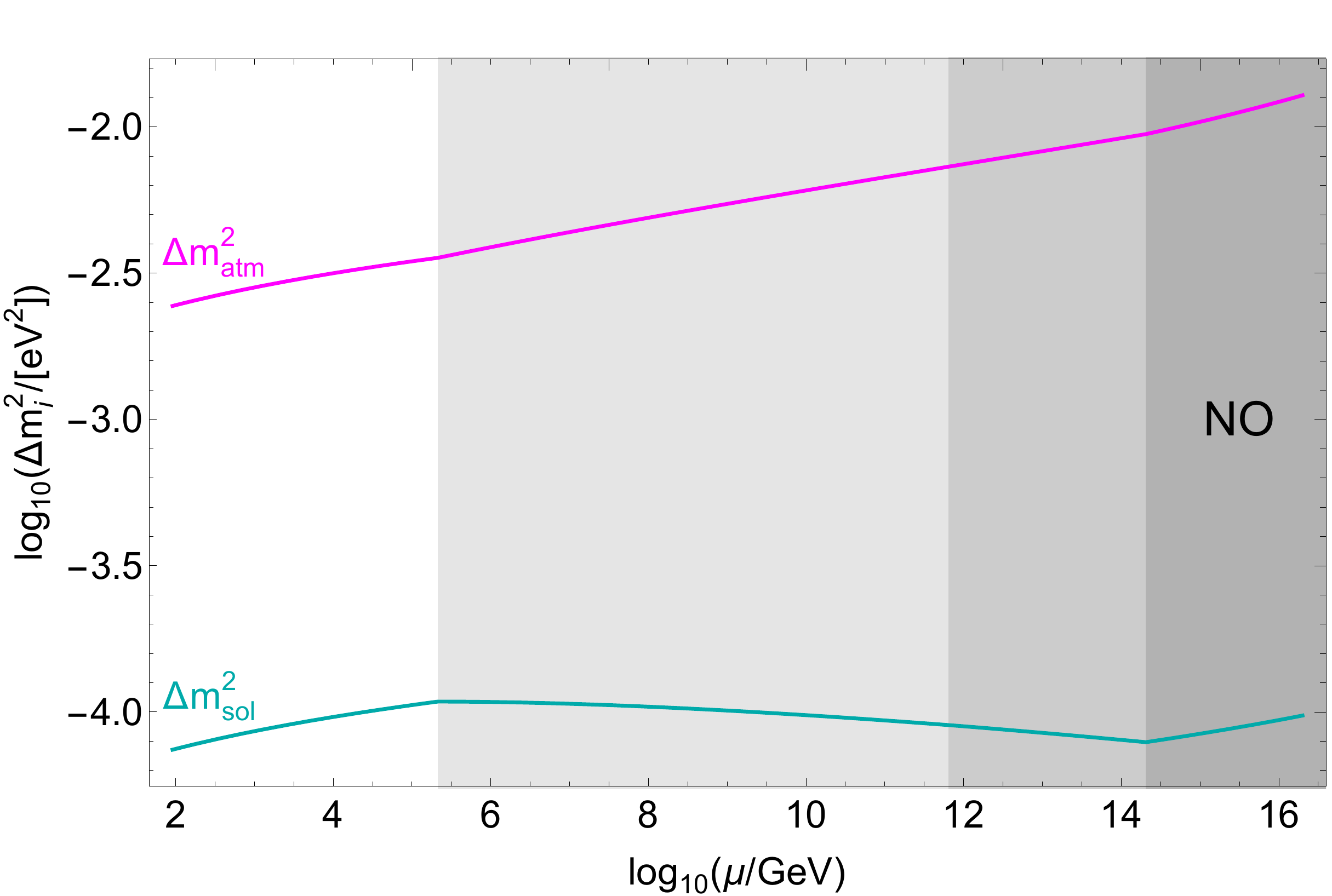}
\includegraphics[width=0.48\textwidth]{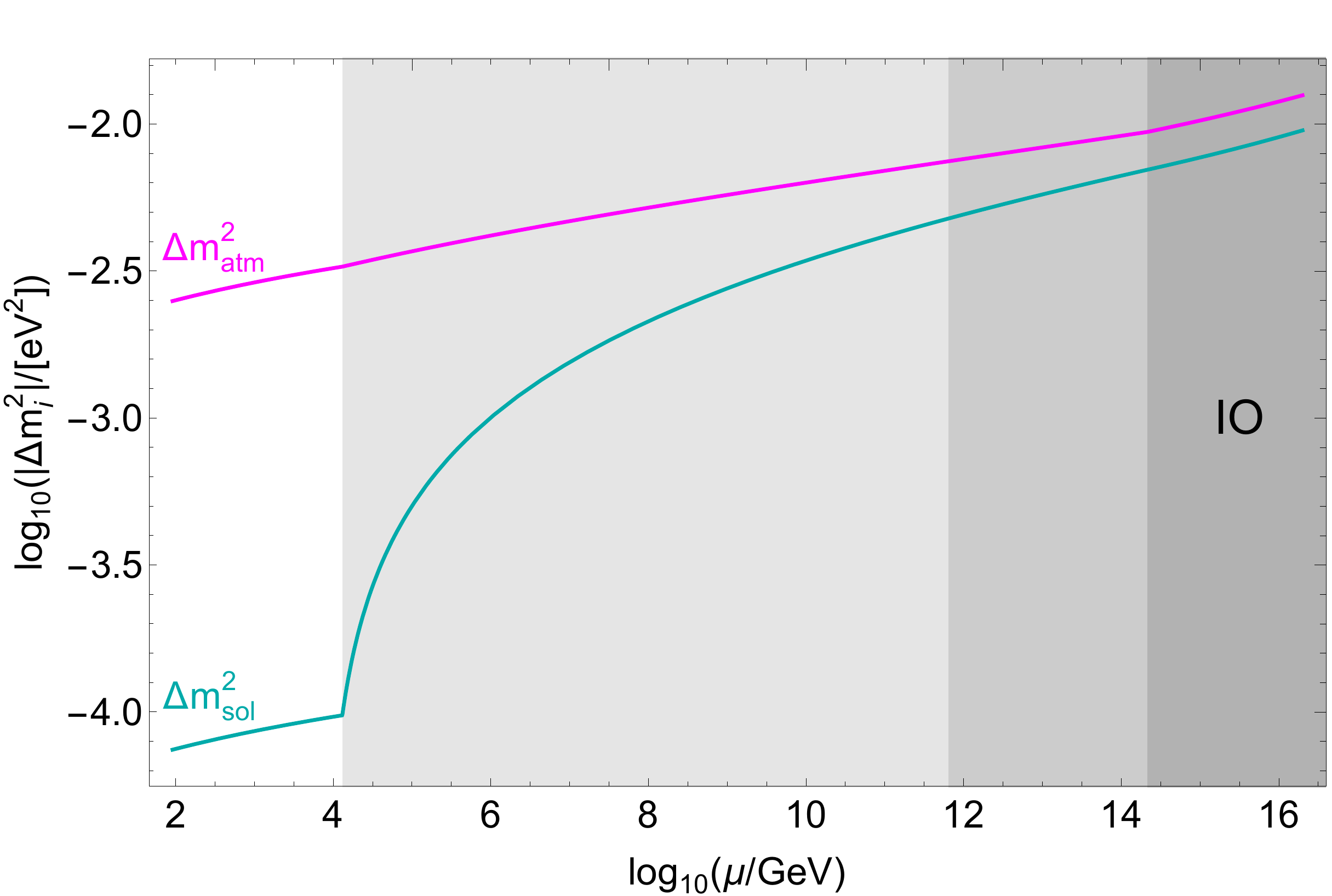}
\\
\includegraphics[width=0.48\textwidth]{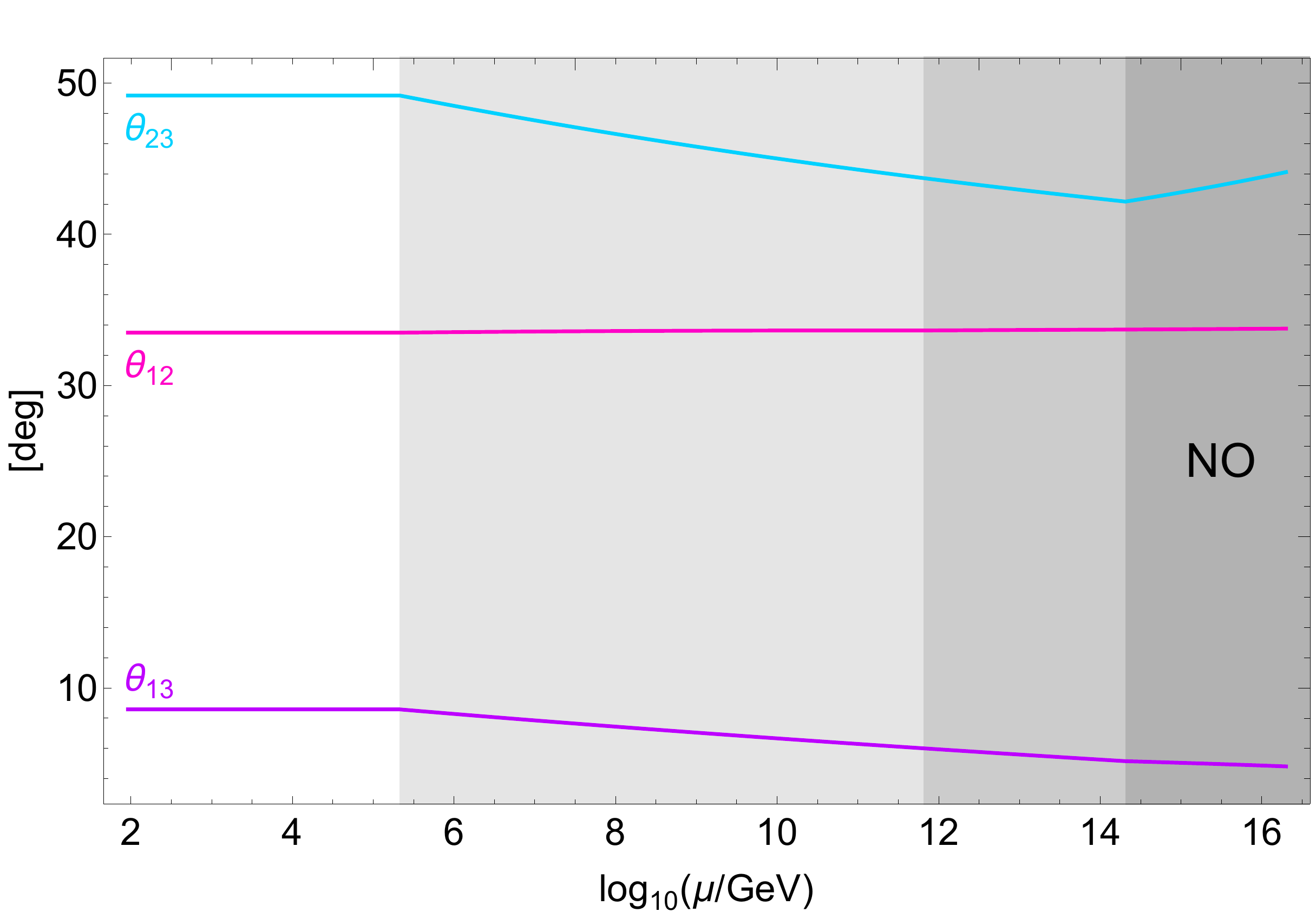}
\includegraphics[width=0.48\textwidth]{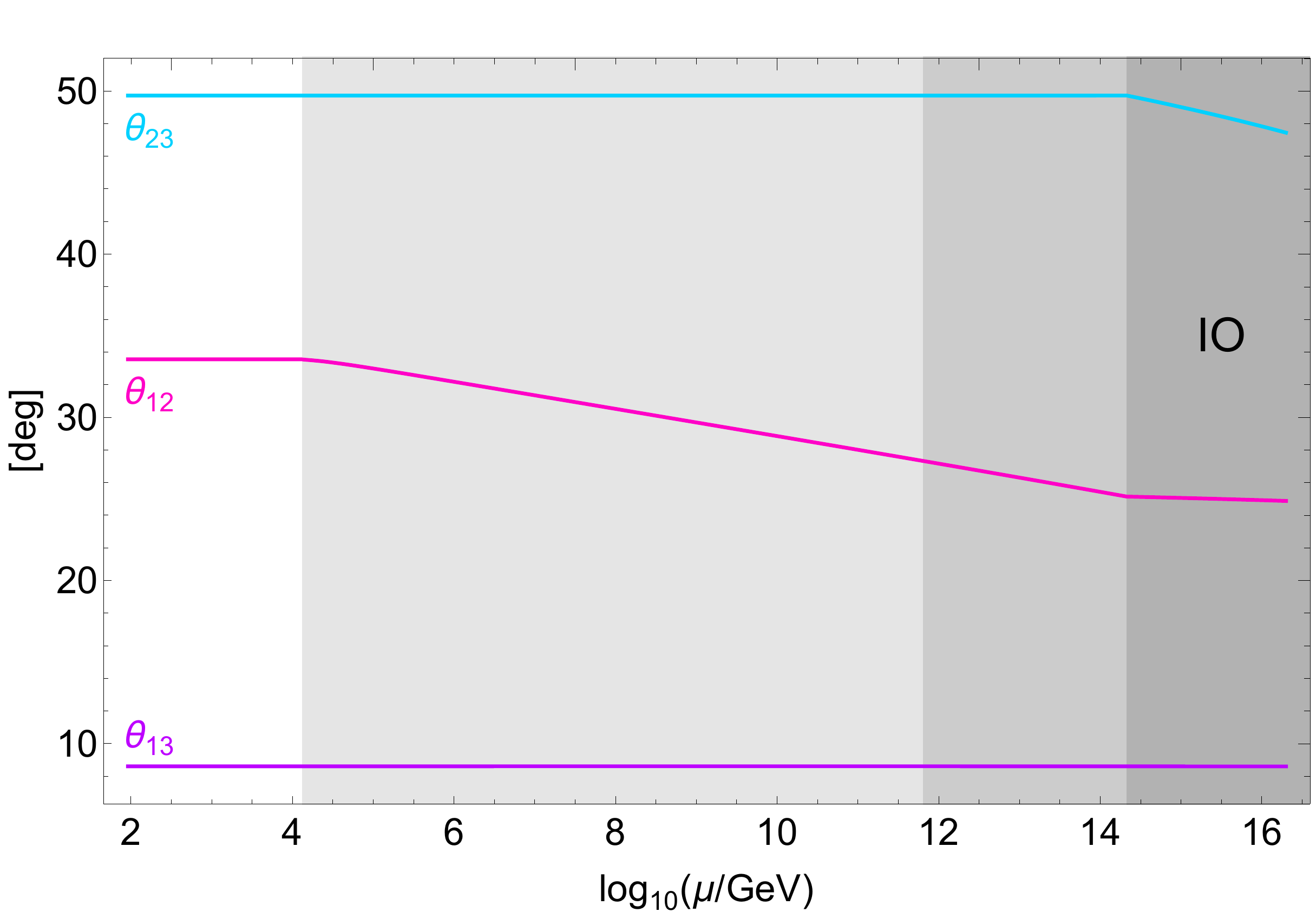}
\caption{ Plots depicting the importance of the RGE running on the tau lepton mass and neutrino mass-squared as well as mixing parameters. Scales, where the heavy right-handed neutrinos $\nu_{R_i}$ decouple from the theory, are presented via different shades of gray.  } \label{fig:RGE}
\end{figure}

As can be seen from  Table~\ref{tab:predictions}, this theory predicts right-handed neutrinos that show a very strong hierarchy, which is a generic feature of this model. To have an understanding of this behavior, let us consider the   best fit parameters given  in Eqs.~\eqref{fit1:NO}-\eqref{fit2:NO} (this behavior, however, is realized for any parameter set that provides a realistic fermion fit). One finds that $(D_{22}, \{D_{33},r_1S_{33}\})/v\sim (10^{-4}, 10^{-3})$ ($r_1S_{33}$ is smaller in order than $D_{22}$). These are the correct orders to reproduce the Yukawas in the down sector, namely, $y_{s,\mu}\sim 10^{-4}$ and $y_{b,\tau}\sim 10^{-2}$ for the second and the third generations, respectively. However, these are too small to provide correct masses to the up-type quarks. Consequently, the charm- and top-quark masses are entirely coming from the matrix $S$. This can be seen by looking at the relevant entries, i.e.,  $(S_{22}, S_{33})/v\sim (10^{-3}, 0.5)$, as needed to replicate $y_c\sim 10^{-3}$ and $y_t\sim 0.5$. On the other hand, $D_{11}/v \sim 10^{-6}$ (in association with other relevant entries in $D$ and $A$ matrices not listed here) reproduces the first generation fermion masses.

Moreover, naively speaking, neutrino mass from the type-I seesaw, with order one coupling demands the seesaw scale to be $M_3\sim 10^{14}$ GeV. Since right-handed neutrino mass matrix is fully determined by the matrix $S$, which is diagonal, $S_{22}/S_{33}\sim y_c/y_t\sim 10^{-3}$ uniquely determines the second lightest right-handed neutrino mass $M_2\sim 10^{-3}M_3\sim 10^{11}$ GeV.

With the entries of order $(D_{11}, \{D_{12,21},A_{12}\}, \{D_{13,3,1},A_{13}\})/v\sim (10^{-6}, 10^{-5}, 10^{-4})$, the structure of the Dirac Yukawa coupling of the neutrino is also hierarchical. As a result, from the type-I seesaw formula Eq.~\eqref{MN}, elements in $M_N$, associated with $M^{-1}_{2,3}$ provide too small (when compared to the 2-3 sector of $M_N$) contributions to  be compatible with neutrino oscillation data. A proper fit to neutrino data requires non-significant entries in the first row and first column of $M_N$ demanding $M_1\ll M_{2,3}$; meaning, all the dominant terms in the first row and first column result from terms that are suppressed only   by  $M_1^{-1}$ in the seesaw formula, which is, however, not true in general, in the 2-3 block. For the fits provided, we find $M_1\sim 10^5\ll M_{2,3}$ GeV to ensure a consistent fit to the neutrino sector for the case of normal mass ordering. On the other hand, for the inverted mass ordering, one must make sure that entries in the first row and column are more dominating compared to the 2-3 sector. This is precisely obtained by ensuring the $M_1$ mass to be slightly smaller than the normal ordering solution. As can be seen from our fits (see Table~\ref{tab:predictions}), $M^\mathrm{IO}_1\sim 10^4$ GeV as required to reproduce inverted neutrino mass ordering, which is one order smaller than  the solution we obtain for normal ordering, $M^\mathrm{NO}_1\sim 10^5$ GeV.     These features in the fermion mass fit as explained above are very distinct from other $SO(10)$ models proposed in the literature.

Before concluding this section, we emphasize on the impact of the RGE running on the neutrino parameters. This is clearly demonstrated in Fig.~\ref{fig:RGE} for the two mass-squared differences and three mixing angles in the neutrino sector. This figure illustrates that a naive fit to the fermion masses and mixing by ignoring the RGEs can lead to significant errors.   Moreover, the top plots  in Fig.~\ref{fig:RGE} show that the running effects on the Tau Yukawa coupling are also very crucial and cannot be neglected.

\section{Symmetry breaking chain and topological defects}\label{sec:breaking}
Since GUT unifies all the known forces present in the SM, it comprises of a larger symmetry group $G$ that contains the SM gauge group ($SU(3)_C\times SU(2)_L\times U(1)_Y \equiv G_{321}\subset G$).  A series of spontaneous summery breaking (SSB) must lead to the SM at  low energies:
\begin{align}
G\to H\to ... \to G_{321} \to G_{31},    \label{PT}
\end{align}
where $G_{31}\equiv SU(3)_C\times U(1)_{em}$, $H\subset G$, and $...$ represents the possibility of multi-step SSB. Subsequent phase transitions in the early Universe, as shown in  Eq.~\eqref{PT}, have direct observational consequences; since various topological defects are produced during cosmological phase transitions  via Kibble mechanism \cite{Kibble:1976sj} that depend on the topology of the vacuum manifold.

Types of topological defects for $G\to H$ are determined by the $n^{th}$ homotopy group $\pi_n (\mathcal{M})$ of the vacuum manifold $\mathcal{M}=G/H$, and defects are formed for $\pi_k (\mathcal{M})\neq 0$~\cite{Vilenkin:2000jqa}. For example,  $\pi_0 (\mathcal{M})\neq 0$, $\pi_1 (\mathcal{M})\neq 0$, and $\pi_2 (\mathcal{M})\neq 0$ lead to domain walls, strings, and monopoles, respectively.  Symmetry breaking patterns that lead to either monopoles or domain walls (or both) are ruled out. If stable topological defects of these types are formed, they would dominate the energy density of the Universe and overclose it, hence must be eliminated to create our observable Universe.

To solve a number of cosmological puzzles (flatness, homogeneity, and isotropy observed in the  CMB data, including the monopole problem), the idea of inflation was put forward~\cite{Guth:1980zm,Albrecht:1982wi,Linde:1981mu,Linde:1983gd}; for a review, see Ref.~\cite{Kolb:1990vq}. If these unwanted topological defects are generated before the inflationary era, they will be diluted away by the Universe's rapid expansion. Production of monopoles is a generic feature of GUTs due to $U(1)\subset G_{321}$; on the contrary, domain walls may or may not appear. To be consistent with observation, in this work, we assume inflation to take place after the last stage of the SSB chain that involves any unwanted defect. Inflation can be successfully implemented with the addition of a GUT singlet, and the detail is not relevant to our discussion here.

Cosmic strings, however, are not ruled out by measurements and may lead to stimulating cosmological signatures. As aforementioned, fascinatingly, pulsar-timing arrays   NANOGrav~\cite{NANOGrav:2020bcs}, PPTA~\cite{Goncharov:2021oub}, EPTA~\cite{Chen:2021rqp}, and IPTA~\cite{Antoniadis:2022pcn}  might
have already seen evidence of a primordial gravitational
wave background that could originate from the last stage spontaneous symmetry breaking in $SO(10)$ GUT. In such a scenario, determining the intermediate symmetry breaking scale from low energy fermion masses and mixings data as well as gravitational
wave signal strongly constrains  the GUT scale, which subsequently has a profound effect on determining the proton lifetime. To accommodate  cosmic strings within a \textit{minimal} (i.e., requiring the least number of Higgses and lowest dimensional representations) framework, we consider the following symmetry breaking chain: 
\begin{align}\label{eq:SSB}
SO(10) 
&\xrightarrow[54_H]{M_X} 
SU(4)_{C}\times SU(2)_{L} \times SU(2)_{R}\times D \equiv G_{422D}
\\
&\xrightarrow[45_H]{M_I} 
SU(3)_{C}\times SU(2)_{L} \times SU(2)_{R}\times U(1)_{B-L} \equiv G_{3221}
\\
&\xrightarrow[126_H]{M_{II}} 
SU(3)_{C}\times SU(2)_{L} \times U(1)_{Y} \equiv G_{321}
\\
&\xrightarrow[10_H+126_H]{M_{EW}} 
SU(3)_{C} \times U(1)_{em} \equiv G_{31}
\end{align}
such that $M_X> M_I>M_{II}>M_{EW}$ (here $M_{X}\equiv M_\mathrm{GUT}$,  $M_{I}\equiv M_\mathrm{PS}$, $M_{II}\equiv M_\mathrm{LR}$).

At the first step, the GUT symmetry is spontaneously broken by the VEV of $54_H$ into Pati-Salam symmetry with the discrete $D$-parity~\cite{Chang:1983fu}. In the next symmetry breaking scale, the VEV of $45_H$ breaks Pati-Salam symmetry into left-right symmetry. At this stage, the $D$-parity is also broken since $45_H$ contains a parity-odd singlet under $G_{3221}$. The remaining symmetry is then  subsequently broken to the Standard Model gauge group via the VEV of $126_H$ Higgs that contains a singlet under $G_{321}$. To be precise, the breaking by $126_H$ leaves a remnant $Z_2$ symmetry (it is the $Z_2$ subgroup of $Z_4$ , the center of SO(10)~\cite{Kibble:1982ae}, which is not broken by tensor representations, regardless of the breaking chain). If this $Z_2$ were absent, embedded strings would appear, which are in general unstable under small perturbations; for a list of breaking chains leading to embedded strings, see, for example, Ref.~\cite{Jeannerot:2003qv}.

Since stable monopoles are formed during the Pati-Salam symmetry breaking,   we assume inflation to take place right after $M_I$ breaking scale that gets rid of all unwanted topological defects. The last symmetry breaking scale (prior to the appearance of the SM gauge group) is carried out by $126_H$ Higgs, and  topologically stable strings with mass
scale related to $M_{II}=v_R$ are formed during this phase transition. Finally,  the SM gauge symmetry is broken  by the VEV of the Higgs doublet, which comes from a combination of the bi-doublets in $10_H$ and $126_H$ dimensional Higgses. For related works on GWs arising from cosmic strings within $SO(10)$ framework, see, for example, Refs.~\cite{Buchmuller:2019gfy,King:2020hyd,King:2021gmj,Masoud:2021prr,Lazarides:2021uxv,Lazarides:2022jgr,Maji:2022jzu,Fu:2022lrn,Lazarides:2022ezc}.

\section{Gravitational waves from cosmic strings}\label{sec:string}
\subsection{Observational signals from strings}
We assume the ideal Nambu-Goto string approximation, where the dominant radiation emission is in the form of GWs~\cite{Vachaspati:1984gt} (for different modeling approaches, see, for example, Ref.~\cite{Auclair:2019wcv}). 
Macroscopic properties of strings are characterized by their energy per unit length $\mu$, known as the string tension parameter, which is given by~\cite{Hindmarsh:2011qj},
\begin{align}
&\mu\sim 2\pi v^2_R B(2\lambda/e^2),
\\
&B(x)\sim 
\begin{cases}
1,\;\;\; x=1.\\
1.04 x^{0.195},\;\;\; 10^{-2}< x \ll 1.
\end{cases}
\end{align}   
With the above, we obtain $G\mu=4.22\times 10^{-38} v^2_R$, where $G=m^{-2}_\textrm{pl}$ denotes the Newton's constant ($m_\textrm{pl}=1.22\times 10^{19}$ GeV). This quantity, $G\mu$,  parametrizes the gravitational interactions of the string. While the existing loops oscillate and emit radiation dominantly to GWs, the expansion of the Universe causes  the long strings to intersect to form new loops.  The power radiated in GWs is given by 
\begin{align}
P_{GW}=\Gamma G \mu^2,
\end{align}
where $\Gamma= 50$~\cite{
Vilenkin:1981bx,Turok:1984cn,Quashnock:1990wv,Blanco-Pillado:2013qja,Blanco-Pillado:2017oxo} is a dimensionless quantity. For Nambu-Goto strings, only large loops give a dominant contribution to GWs, which has the initial length of $\ell_i=\alpha t_i$, where $\alpha$ characterizes the loop size at the time of formation, and numerical simulation typically corresponds to $\alpha= 0.1$~\cite{Blanco-Pillado:2013qja,Auclair:2019wcv}.  Due to energy loss, their lengths decrease with time $\ell=\alpha t_i-\Gamma G \mu (t-t_i)$, until loops disappear completely.  The frequency associated to the GW from the total energy loss from a loop is $f_k=2 k/\ell$, where $k=1,2,3,...$ is the mode number. Then the relic GW background from a cosmic string network has the following form~\cite{Blanco-Pillado:2017oxo,Cui:2018rwi}: 
\begin{align}
&\Omega_{GW}(f)=\sum_k \Omega_{GW}^{(k)}(f), \label{gw}
\end{align}
where the sum over all modes are performed, and the explicit expression  of $\Omega_{GW}^{(k)}(f)$ is given by 
\begin{align}
&\Omega_{GW}^{(k)}(f)=\frac{1}{\rho_c}\frac{2k}{f}\frac{\mathcal{F}_\alpha \Gamma^{(k)}G \mu^2}{\alpha(\alpha+\Gamma G \mu)}\int^{t_0}_{t_F}d\widetilde{t} \frac{c_{eff}(t^{(k)}_i)}{{t^{(k)}_i}^4}\left( \frac{a(\widetilde{t})}{a(t_0)}\right)^5  \left(\frac{a(t^{(k)}_i)}{a(\widetilde{t})}\right)^3 \Theta\left(t^{(k)}_i-t_F\right), 
\end{align}
here $\Gamma^{(k)}=\Gamma k^{-4/3}/3.6$ and the integral is taken over the GW emission time $\widetilde{t}$, $t_F$ is the time of the formation of cosmic string network, and the formation time of loops contributing with $k$-th mode number is defined as
\begin{align}
t^{(k)}_i(\widetilde{t},f)=\frac{1}{\alpha+\Gamma G \mu} \left( \frac{2 k}{f} \frac{a(\widetilde{t})}{a(t_0)} +\Gamma G \mu \widetilde{t}  \right).    
\end{align}
Furthermore,  $\rho_c=3  H^2_0/(8\pi G)$ is the critical energy density of the Universe,  $\mathcal{F}_\alpha= 0.1$~\cite{Blanco-Pillado:2013qja}, and $c_{eff}= 5.7$ (0.5) for radiation (matter) dominance are determined numerically.

\begin{figure}[t!]
\centering
\includegraphics[width=0.85\textwidth]{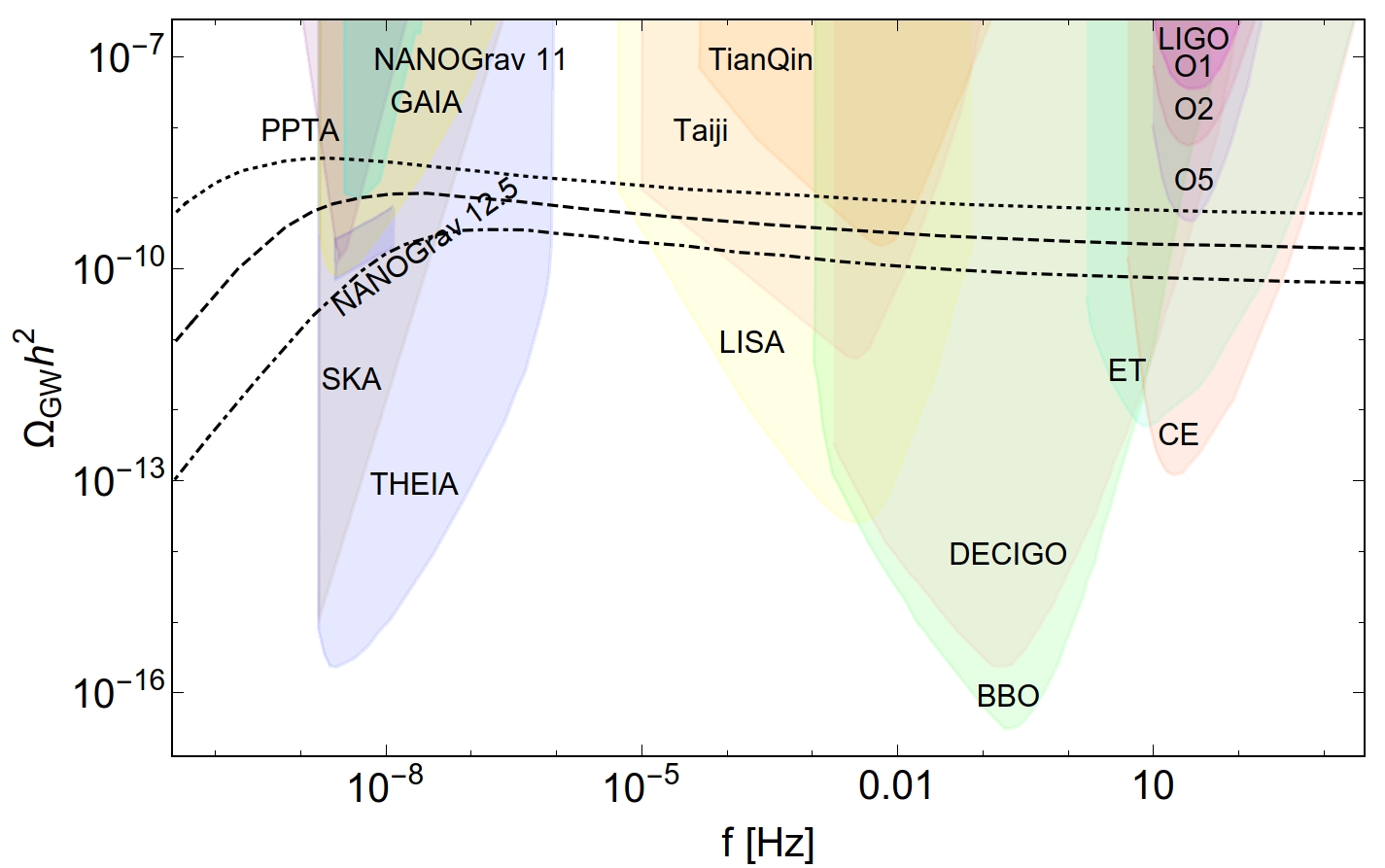}
\caption{ Gravitational wave signals from stable cosmic string networks as a function of frequency observed today together with the corresponding sensitivities of
the indicated observatories. See text for details.  } \label{fig:GW}
\end{figure}

Gravitational wave signals from stable cosmic string networks as a function of frequency observed today together with the  sensitivities of
various observatories are presented in Fig.~\ref{fig:GW} for three chosen values of $G\mu$.    CMB measurements directly provide constraints on string tension $G\mu < 1.1\times 10^{-7}$ for Nambu-Goto strings~\cite{Planck:2015fie,Charnock:2016nzm}, which corresponds to $M_{II} \lesssim 2\times 10^{15}$ GeV, hence, GUT scale appearance of strings are ruled out (if no fine-turning is done, then the current experimental bounds on proton lifetime requires the GUT scale to be $M_\mathrm{GUT}\gtrsim 5\times 10^{15}$ GeV). Recently, LIGO updated their O3 data~\cite{LIGOScientific:2021nrg} that rules out $G\mu\gtrsim 10^{-8}$ in the $f\sim \mathcal{O}(10)$ Hz regime, providing $M_{II} \lesssim 5\times 10^{14}$ GeV.   Moreover, non-observation of SGWs in the $f\sim$nHz region  by EPTA \cite{Lentati:2015qwp} and 11 year data of NANOGrav \cite{Arzoumanian:2018saf} rule out string formation up to almost $10^{14}$ GeV.

On the other hand, 12.5 years of NANOGrav's data  indicates SGWs for string tension in the range $G\mu\sim [4\times 10^{-11}-10^{-10}]$ at $68\%$ confidence level (CL), while $95\%$ CL corresponds to the range $G\mu\sim [2\times 10^{-11}-3\times 10^{-10}]$  \cite{Ellis:2020ena}. As seen from Fig.~\ref{fig:GW}, since GW spectrum of cosmic strings is very flat over a wide frequency range, the current hints of SGW backgrounds at PTAs for sure can be confirmed by the measurement of various upcoming experiments.    Frequencies in the window $10^{-5}-10^3$ Hz will be probed by ground-based interferometers: Einstein Telescope (ET)~\cite{Sathyaprakash:2012jk} and Cosmic Explorer (CE)~\cite{Evans:2016mbw}; space-based laser interferometers  Laser Interferometer Space Antenna (LISA)~\cite{Audley:2017drz}, Taiji~\cite{Guo:2018npi}, TianQin~\cite{Luo:2015ght}, BBO~\cite{Corbin:2005ny}, and DECIGO~\cite{Seto:2001qf}); atomic interferometers MAGIS~\cite{Graham:2017pmn}, AEDGE~\cite{Bertoldi:2019tck}, AION~\cite{Badurina:2019hst}. Whereas  pulsar timing arrays such as EPTA  and NANOGrav are currently covering the nHz regime. Future pulsar timing array SKA~\cite{Janssen:2014dka} will also cover a similar frequency regime which is sensitive to much smaller GW energy densities. Furthermore, large-scale surveys of stars such as Gaia~\cite{Gaia:2018ydn} and the proposed upgrade, THEIA~\cite{Theia:2017xtk} will be covering $1-100$ nHz regime, where the latter is sensitive to GW energy densities as low as $10^{-16}$. These experiments are sensitive to GWs arising from cosmic strings with breaking scales $v_R\gtrsim 10^{9}$ GeV. It is remarkable that these GW experiments will probe GUT scale physics and can rule out various symmetry breaking chains; such large energy scales, however, are impossible to probe in the colliders.

\begin{figure}[t!]
\centering
\includegraphics[width=0.8\textwidth]{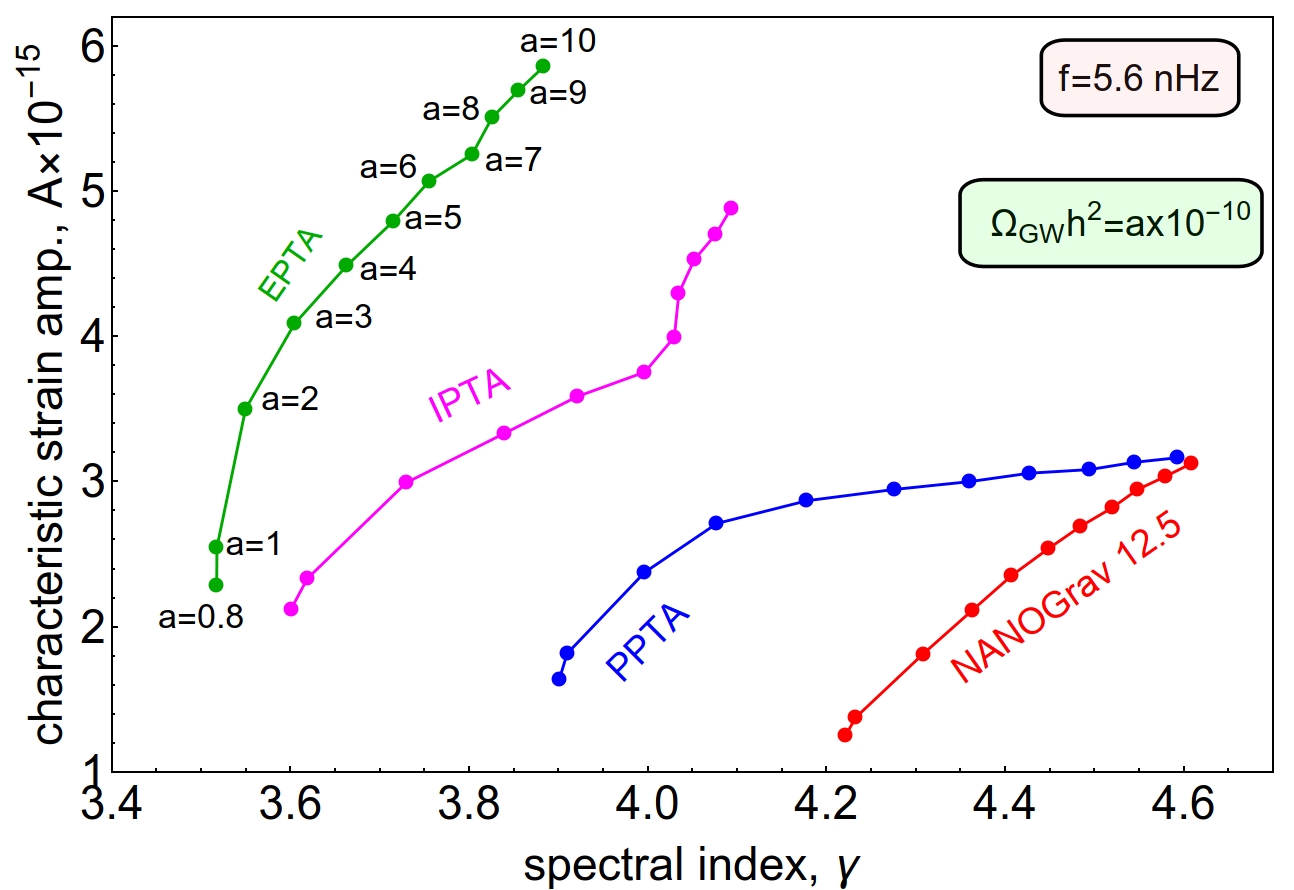}
\caption{  Best fit values of the  parameters $(A,\gamma)$  for the various PTAs with a fixed frequency, $f=5.6$ nHz. Good fits (i.e., $\sum\chi^2<2$) are obtained for $a\sim[1,6]$.  See text for details.  } \label{fig:PTA}
\end{figure}

The observation of GWs at the PTAs
NANOGrav~\cite{NANOGrav:2020bcs}, PPTA~\cite{Goncharov:2021oub}, EPTA~\cite{Chen:2021rqp}, and IPTA~\cite{Antoniadis:2022pcn}  hint towards gravitational wave signals with the energy density given by~\cite{Fu:2022lrn},
\begin{align}
\left(\Omega_{\rm GW}(f) h^2 \right)_{\rm PTA} \approx 2.02\cdot 10^{-10} \left(\frac{A}{10^{-15}}\right)^2 \times \left( \frac{f}{f_{\rm yr}} \right)^{5-\gamma},    
\end{align}
where $f_{\rm yr}=1 yr^{-1}$. $A$ is the characteristic strain amplitude and $\gamma$ is the spectral index.  In fact, NANOGrav~\cite{NANOGrav:2020bcs} finds strong evidence of a stochastic process modeled as a power-law, $f^{-2/3}$, with common amplitude and spectral slope across pulsars. This power-law spectrum of the characteristic GW strain corresponds to $\gamma=13/3$, which is motivated by searches for mergers of supermassive black hole binaries. Within this fiducial model, the amplitude parameter $A$ that relates the correlation between
pulsars has a median $1.92\times 10^{-15}$ at $95\%$ confidence level. Without fixing the value of $\gamma$, other   pulsar timing arrays also find the amplitude of similar values in the same frequency range. Fitted values of these  pulsar timing array experiments are summarized in Table~\ref{tab:A-gamma}. 
\begin{table}
\centering
\begin{tabular}{|c |c |c |} 
 \hline
 Experiment&$A\times 10^{15}$&$\gamma$ \\ [0.5ex] 
 \hline\hline
 NANOGrav & $1.92_{-0.55}^{+0.75}$ & $4.333$ \\ 
 \hline
 PPTA & $2.82_{-1.16}^{+0.73}$ & $4.11_{-0.41}^{+0.52}$ \\
 \hline
 EPTA & $5.13_{-2.73}^{+4.20}$ & $3.78_{-0.59}^{+0.69}$ \\
 \hline
 IPTA & $3.8_{-2.5}^{+6.3}$ & $4.0_{-0.9}^{+0.9}$ \\ [1ex] 
 \hline
\end{tabular}
\caption{Experimentally allowed values of the characteristic strain amplitude and the spectral index at $95\%$ CL.}\label{tab:A-gamma}
\end{table}
Intriguingly, these data from various pulsar timing array experiments strongly favor stochastic gravitational-wave background in the nHz range with energy density
\begin{align}
\left(\Omega_{\rm GW} h^2 \right)_{\rm PTA}\sim 10^{-10},    
\end{align}
which is in the ballpark, as expected from $SO(10)$ GUTs.

\begin{figure}[t!]
\centering
\includegraphics[width=0.8\textwidth]{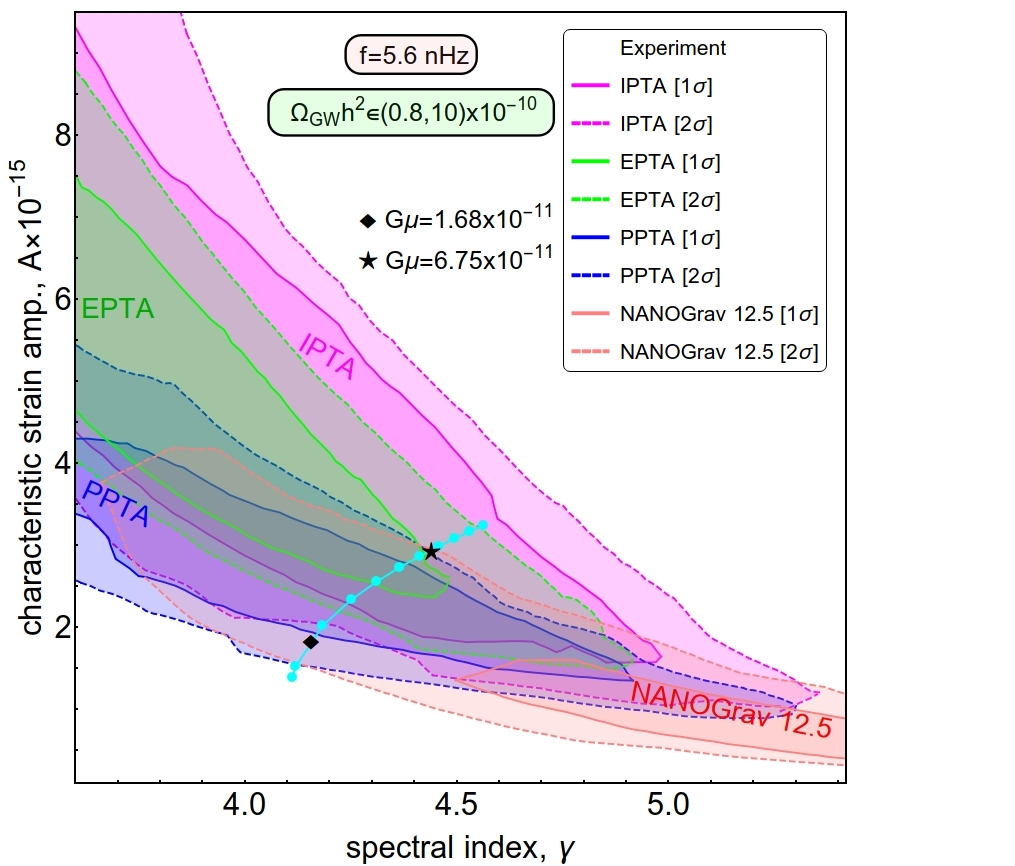}
\caption{ A combined fit (cyan curve) to NANOGrav, PPTA, EPTA, and IPTA data in the $(A,\gamma)$ plane. The diamond and the star shaped points are the two chosen benchmark scenarios of the proposed model. See text for details.   } \label{fig:PTAreg}
\end{figure}

By using the experimental data shown in Table~\ref{tab:A-gamma}, we present the best fit values of the  parameters $(A,\gamma)$  in Fig.~\ref{fig:PTA} for the separate PTAs by fixing $\Omega_{\rm GW}h^2$ in the range $8\times 10^{-11}-10^{-9}$.  For this analysis, we have fixed the frequency at $f=5.6$ nHz, which is within the frequency range $f\in (2.5,12)$ nHz considered by NANOGrav~\cite{NANOGrav:2020bcs} (see also Ref.~\cite{Ellis:2020ena}). For our numerical study, we allow a $5\%$ fluctuations of $\gamma_\mathrm{NANOGrav}=13/3$.

To illustrate that the proposed model can simultaneously be consistent with these experimental measurements,  we perform a combined fit to these data. In doing so, first, we  symmetrize the experimental errors on the two observables, i.e., $A^{(k)}\pm \sigma^{(k)}_A$ and $\gamma^{(k)}\pm \sigma^{(k)}_\gamma$; $k=$NANOGrav, IPTA, PPTA, and EPTA.  Since these experimental values of the same physical quantity are  uncorrelated, we get the   best estimate via~\cite{Fabjan:2020wnt}
\begin{align}
&A^{\rm comb.}=\frac{\sum_{k} \frac{A^{(k)}}{\left(\sigma^{(k)}_A\right)^2 }}{\sum_k \frac{1}{\left(\sigma^{(k)}_A\right)^2}}, \;\;\; \sigma^{\rm comb.}_A=    \frac{1}{\sqrt{\sum_k \frac{1}{\left(\sigma^{(k)}_A\right)^2}}},
\\
&\gamma^{\rm comb.}=\frac{\sum_{k} \frac{\gamma^{(k)}}{\left(\sigma^{(k)}_\gamma\right)^2 }}{\sum_k \frac{1}{\left(\sigma^{(k)}_\gamma\right)^2}}, \;\;\; \sigma^{\rm comb.}_\gamma=    \frac{1}{\sqrt{\sum_k \frac{1}{\left(\sigma^{(k)}_\gamma\right)^2}}}.
\end{align}
The result of our numerical fit is presented in  Fig.~\ref{fig:PTAreg}. This figure shows that an excellent fit to the data from PTAs can be obtained for $\Omega_\mathrm{GW}h^2\sim 10^{-10}$. We find that for $f=5.6$ nHz, in the range of $\Omega_\mathrm{GW}h^2\in (2,6)\times 10^{-10}$, all these PTAs observations can be satisfied within $2\sigma$ CL. This range roughly corresponds to  $G\mu\in (4.9,6.9)\times 10^{-11}$. In other words, the seesaw scale from the observation of GWs is highly restricted within $v_R\in (3.4,4.1)\times 10^{13}$ GeV.

\subsection{Fit to fermion mass with fixed intermediate scale} 
As discussed in the previous sub-section, data from PTAs prefers $v_R\sim 4\times 10^{13}$ GeV.  On the contrary, as we have seen in Sec.~\ref{sec:yukawa} (see Table~\ref{tab:predictions}),  this theory, for both the NO and IO scenarios, prefers $M_3\sim 2\times 10^{14}$ GeV, which is in direct conflict (assuming perturbative Yukawa couplings) with observations of pulsar-timing arrays.   To lower the seesaw scale from the best fit value, we perform  an extensive numerical analysis and find that  for $M_3$ as low as $M_3\sim 7\times 10^{12}$ GeV,  $\chi^2\sim \mathcal{O}(10)$, which arises since the up-quark mass cannot be fitted to its observed value.  This corresponds to more than $3\sigma$ discrepancy in the up-quark mass measurement, which is unacceptable.   The $\Delta\chi^2$ contribution to the up-quark mass reduces to about $2\sigma$ for  $M_3=2\times 10^{13}$ GeV, which is in the acceptable range. Starting from this value, an excellent fit to all observables can be found up to about $M_3\sim 10^{15}$ GeV. Above this scale, neutrino masses cannot be correctly incorporated.  Hence, the low energy observables are consistent with experimental measurements provided that $M_3= [2\times 10^{13}-10^{15}]$ GeV.

To be consistent with both low energy fermion masses and mixings data as well as results of pulsar-timing arrays,  as a benchmark, we take $M_3=4\times 10^{13}$ in what follows, which corresponds to  $v_R\in(2,8)\times 10^{13}$ GeV depending on the value of the chosen Yukawa coupling, see Eq.~\eqref{vR:range}. Two benchmark values, $v_R=2\times 10^{13}$ and $v_R=4\times 10^{13}$ GeV corresponding to $G\mu=1.688\times 10^{-11}$ (diamond shaped point) and $G\mu=6.754\times 10^{-11}$ (star shaped point) are highlighted in Fig.~\ref{fig:PTAreg}.  

Other than fixing the seesaw scale, 
the fitting procedure performed in this section is identical to the prescriptions outlined in Sec.~\ref{sec:yukawa}. This analysis is performed by assuming normal ordering; however, repeating it for inverted ordering is expected to result in similar conclusions.  The best fit result of the observables is presented in Table~\ref{result:fix}. This solution corresponds to a total $\chi^2=0.52$, which receives  the largest contribution from the up-quark mass corresponding to $\Delta \chi^2\sim 0.34$. The best fit parameters are presented in Eqs.~\eqref{fit1:fix}-\eqref{fit2:fix}.  Proton decay branching fraction predicted from this best fit is summarized in Table~\ref{tab:BRpredictions}.

\FloatBarrier
\begin{table}[th!]
\centering
\footnotesize
\resizebox{0.99\textwidth}{!}{
\begin{tabular}{|c|c||c||c|}
\hline

\textbf{Observables} & \textbf{Best fit ($M_Z$ scale)} & \textbf{Observables} & \textbf{Best fit ($M_Z$ scale)}  \\ 
\hline \hline

\rowcolor{red!10}$y_u/10^{-6}$&5.33&  \cellcolor{orange!10}$\theta_{12}^\textrm{CKM}/10^{-2}$&\cellcolor{orange!10}$22.737$
\\  
\rowcolor{red!10}$y_c/10^{-3}$&3.60& \cellcolor{orange!10}$\theta_{23}^\textrm{CKM}/10^{-2}$&\cellcolor{orange!10}4.208 \\ 
\rowcolor{red!10}$y_t$&0.986& \cellcolor{orange!10}$\theta_{13}^\textrm{CKM}/10^{-3}$&\cellcolor{orange!10}3.63 \\ 

\rowcolor{yellow!10}$y_d/10^{-5}$&1.698& \cellcolor{orange!10}$\delta^\textrm{CKM}$&\cellcolor{orange!10}1.207 \\ 
\rowcolor{yellow!10}$y_s/10^{-4}$&3.126& \cellcolor{green!10}$\Delta m^2_{21}/10^{-5} (eV^2)$&\cellcolor{green!10}7.403 \\ 
\rowcolor{yellow!10}$y_b/10^{-2}$&1.639& \cellcolor{green!10}$\Delta m^2_{31}/10^{-3} (eV^2)$ &\cellcolor{green!10}2.517\\ 

\rowcolor{cyan!10}$y_e/10^{-6}$&2.7942& \cellcolor{blue!10}$\sin^2 \theta_{12}$ &\cellcolor{blue!10}0.3028 \\ 
\rowcolor{cyan!10}$y_\mu/10^{-4}$&5.8961& \cellcolor{blue!10}$\sin^2 \theta_{23}$   &\cellcolor{blue!10}0.5723\\ 
\rowcolor{cyan!10}$y_\tau/10^{-2}$&1.0028& \cellcolor{blue!10}$\sin^2 \theta_{13}$ &\cellcolor{blue!10}0.02233\\ \hline

\end{tabular}
}
\caption{Best fit values of the observables by restricting the intermediate symmetry breaking scale. In particular, we fix $M_3=4\times 10^{13}$ GeV that results in $\chi^2=0.52$, hence representing an excellent agreement with measurements.  See text for details.   }\label{result:fix}
\end{table}

\begin{table}[th!]
\centering
{\footnotesize
\resizebox{0.7\textwidth}{!}{
    \begin{tabular}{|c|c|}
    \hline
       {\bf $p$ decay modes}  & {\bf BR[$\%$] predictions}  \\
\hline \hline

$\left(p\rightarrow e^+ \pi^0,p\rightarrow e^+ K^0,p\rightarrow e^+ \eta \right)$  & $\left( 32.4, 0.52, 0.03  \right)$       
 \\[3pt]
\hline     

$\left(p\rightarrow \mu^+ \pi^0, p\rightarrow \mu^+ K^0,p\rightarrow \mu^+ \eta \right)$  & $\left( 0.42, 4.3, 0.002  \right)$          
 \\[3pt]
\hline     

$\left(p\rightarrow \overline{\nu} \pi^+,p\rightarrow \overline{\nu} K^+ \right)$  & $\left(61.5,  0.81 \right)$     
 \\[3pt]
\hline     
    \end{tabular}
    }
    \caption{Proton decay branching fraction predictions of the best fit with the fixed seesaw scale, $M_3=4\times 10^{13}$ GeV.  }
    \label{tab:BRpredictions}
    }
\end{table}

\begin{figure}[th!]
\centering
\includegraphics[width=0.8\textwidth]{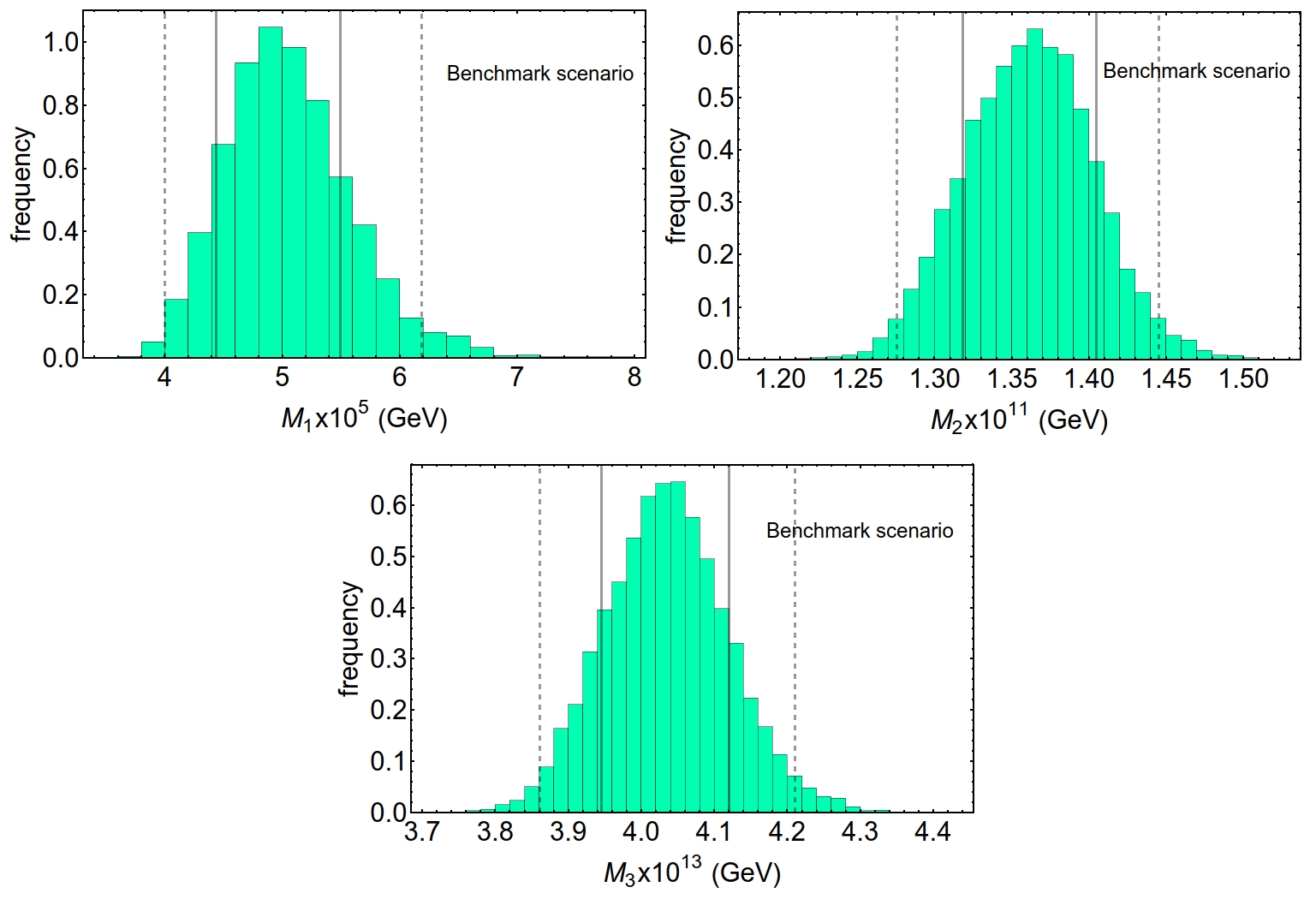}
\caption{ Right-handed neutrino masses in the benchmark scenario resulting from MCMC analysis. The vertical solid (dashed) line corresponds to  $1\sigma$ ($2\sigma$) highest posterior Density (HPD) for the masses $M_i$.} \label{fig:MR}
\end{figure}

\begin{figure}[th!]
\centering
\includegraphics[width=0.55\textwidth]{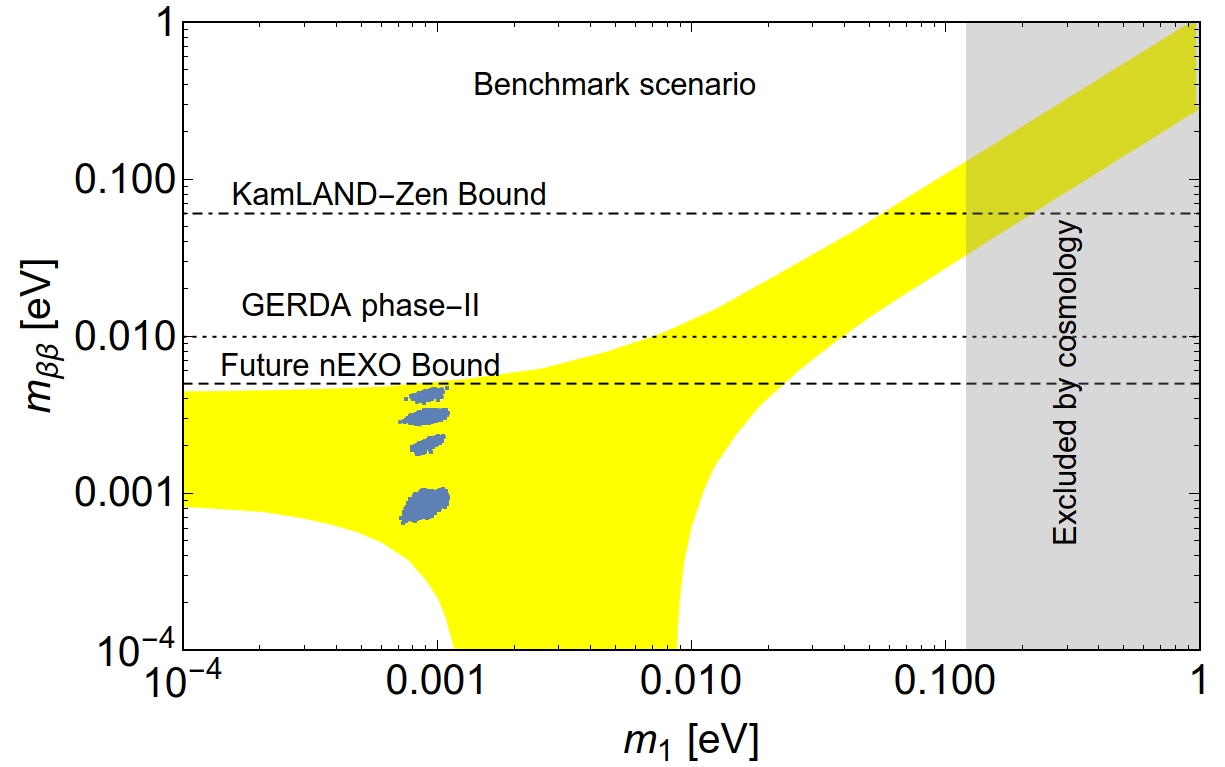}
\caption{ Neutrinoless double beta decay in the benchmark scenario resulting from MCMC analysis. Cosmological exclusion bound (on the sum of the neutrino masses) is taken from Ref.~\cite{ParticleDataGroup:2020ssz}.
} \label{fig:mBETA}
\end{figure}

\begin{figure}[th!]
\centering
\includegraphics[width=0.55\textwidth]{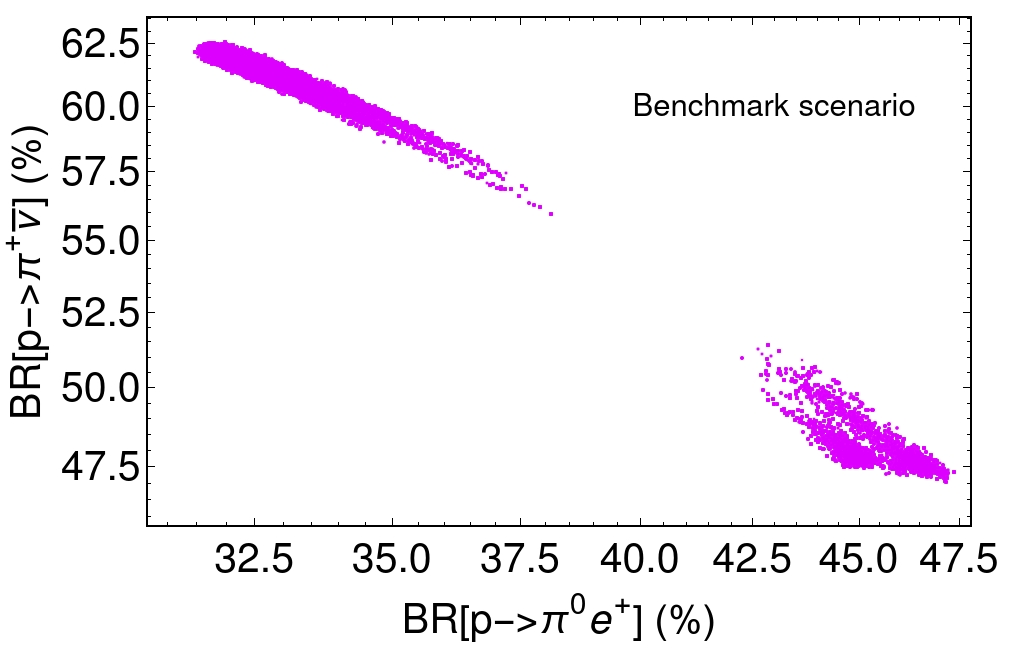}
\caption{ Proton decay branching ratios of the two most dominating channels in the benchmark scenario resulting from MCMC analysis.  } \label{fig:pDECAY}
\end{figure}

\begin{figure}[th!]
\centering
\includegraphics[width=0.55\textwidth]{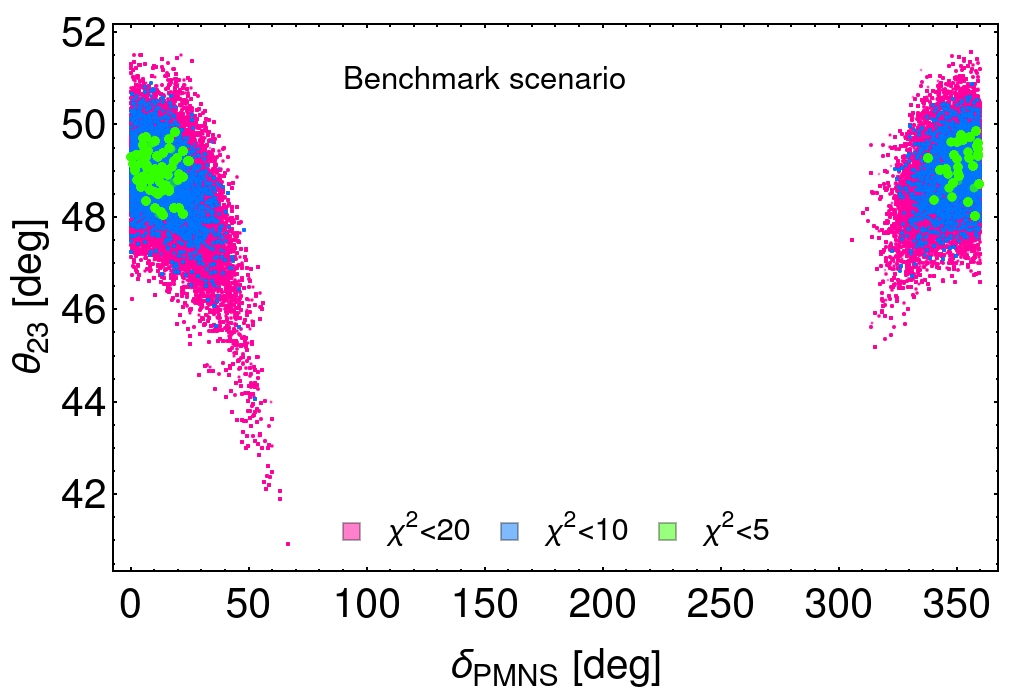}
\caption{ Correlation between the Dirac CP phase and $\theta_{23}$ in the neutrino sector  in the benchmark scenario resulting from MCMC analysis.  } \label{fig:PMNS}
\end{figure}

For this benchmark scenario, after finding the best fit,  we perform a MCMC analysis  using an adaptive Metropolis-Hastings algorithm, the results of which are  presented in Figs.~\ref{fig:MR}, \ref{fig:mBETA}, \ref{fig:pDECAY}, and \ref{fig:PMNS}  for selected physical quantities.  Fig.~\ref{fig:MR} shows the Histogram plot of the right-handed neutrino masses $M_i$ with their associated  $1\sigma$ and $2\sigma$ highest posterior Density (HPD) intervals. In Fig.~\ref{fig:mBETA}, we present the  neutrinoless double beta decay parameter along with sensitives from a number of experiments. This includes current bound from KamLAND-Zen~\cite{KamLAND-Zen:2016pfg} and future sensitives of next-generation experiments such as GERDA Phase II~\cite{GERDA:2019cav} and  nEXO~\cite{nEXO:2021ujk}. Fig.~\ref{fig:pDECAY} demonstrates the interrelationship between the two most dominating proton decay modes in this theory. Finally, Fig.~\ref{fig:PMNS} depicts the correlation between the 2-3 mixing angle and the CP violating Dirac phase in the neutrino sector, where the latter is yet to be measured in the experiments.  As mentioned above, in our analysis, we do not include this phase in the $\chi^2$-function.

\section{Proton decay and Gauge coupling unification}\label{sec:PDunification}
Here, we discuss the constraining arising from proton decay and gauge coupling unification in the proposed model. 

\subsection{Proton decay}
Nucleon decay is the smoking gun signal of grand unification (for a recent review, see Ref.~\cite{Dev:2022jbf}). Proton decay in nonsupersymmetric grand unified models is dominated by the exchange of the superheavy gauge bosons, interactions of  which violate baryon number conservation.  Since these gauge bosons have masses at the GUT scale, observation of proton decay would probe the grand unification scale. In our model, the GUT scale is intrinsically connected to the intermediate symmetry breaking scales. As we will show, there is a non-trivial correlation between the proton lifetime and the observation of gravitational waves within the proposed model.

The superheavy gauge bosons lead to proton decays of the form:  $p\to M + \bar{l}$, where the meson $M$ can be $\pi^0$, $\pi^{+}$, $K^0$, $K^{+}, \eta$ and the lepton can be $l= e, \mu,\nu$ \cite{Machacek:1979tx}. The partial decay width for such nucleon decay process is given by the following formula: 
\begin{align}
\Gamma (p\to M + \bar{l}) = \frac{m_p}{32\pi}\left[1-\left(\frac{m_{M}}{m_p}\right)^2\right]^2 A_L^2 \bigg|\sum_n A_{Sn} W_n F_0^n(p\to M)\bigg|^2 .
\end{align}
In the above equation, $m_p$, and $ m_{M}$ are the masses of the proton and Mesons, respectively.  Furthermore, $W_n$ are  the Wilson coefficients of the operators that give rise to a particular decay channel $p\to M + \bar{l}$; $A_{Sn}$'s are the short-range enhancement factors (to be discussed shortly),  and $F_0^n= \langle M \rvert {(q q')}_n q''_L\lvert p \rangle$ is the relevant form factor. Here, $q$, $q'$ and $q''$ are the light quarks, namely, $ u,  d,  s$.

The most important proton decay channel within our framework corresponds to $p\to e^+ \pi^0$, which is experimentally the most constraining one. The current experimental bound from Super-Kamiokande  provides $\tau_p(p\to e^+ \pi^0)> 2.4\times 10^{34}$ yrs.~\cite{Super-Kamiokande:2020wjk}. The expression of the lifetime of this process is then given by \cite{FileviezPerez:2004hn,Nath:2006ut}:
\begin{align}
\tau_p = & \Bigg[ \frac{m_p}{32\pi}\left(1-\frac{m_{\pi^0}^2}{m_p^2}\right)^2 A_L^2 \frac{g_\mathrm{GUT}^4}{4 M_X^4}  \times  
\nonumber \\
& 
\bigg\{ A_{SR}^2 \left| \left[  V_1^{11} V_3^{11}+ \left(V_4V^\dagger_{UD}\right)^{11} \left( V_1 V_{UD} V_4^\dagger V_3\right)^{11} \right]  \langle \pi^0 \rvert (ud)_R u_L\lvert p \rangle \right|^2 
\nonumber \\
& 
+ A_{SL}^2 \left|    \left[ V_1^{11} V_2^{11} +\left( V_1 V_{UD}\right)^{11}\left(V_2 V^\dagger_{UD}\right)^{11} \right]\   \langle \pi^0 \rvert (ud)_L u_L\lvert p \rangle \right|^2 \bigg\} \Bigg]^{-1} ,
\end{align}
where, $M_X$ is the mass of the superheavy gauge boson, and $A_{SL}$ and $A_{SR}$ are the short-range enhancement factors associated with the left-handed $\mathcal{O}^{d=6}_L\left( e^C, d\right)$ and right-handed $\mathcal{O}^{d=6}_R\left( d^C, e \right)$ operators, respectively. Moreover, in our calculation, we use the following values of the matrix elements \cite{Aoki:2017puj}: 
\begin{align*}
\langle \pi^0 \rvert (ud)_R u_L\lvert p \rangle = -0.131, \ \ \, \, \langle \pi^0 \rvert (ud)_L u_L\lvert p \rangle = 0.134 .
\end{align*}
The rest of the matrix elements required to compute proton decay to other channels can also be found in Ref.~\cite{Aoki:2017puj}. The mixing matrices  $V_1, V_2$ etc are defined as follows:
 $V_1= U_R^T U_L$, $V_2=E_R^TD_L$,
$V_3=D_R^TE_L$, $V_4=D_R^T D_L$, $V_{UD}=U_L^{\dagger}D_L$,
$V_{EN}=E_L^{\dagger}N_L$ and $U_{EN}= E^T_R N_L$,
where $U,D,E$ define the  diagonalizing matrices given by
\vspace{-2pt}
\begin{eqnarray}
U^{\dagger}_R \ M_U \ U_L &=& M_U^{diag} \\
D^{\dagger}_R \ M_D \ D_L &=& M_D^{diag} \\
E^{\dagger}_R \ M_E \ E_L &=& M_E^{diag} \\
N^{\dagger}_R \ M_N \ N_L &=& M_N^{diag} .
\end{eqnarray}
with $N_R=N^*_L$. 

Finally, the long distant enhancement factor $A_L$ \cite{Nihei:1994tx} roughly takes the value $A_L\approx 1.2$, and the short range renormalization factors $A_S$ are  written as follows  \cite{Buras:1977yy,Goldman:1980ah,Caswell:1982fx,Ibanez:1984ni}: 
\begin{align} 
A_S = \prod_{j}^{M_Z\leq M_{j}\leq M_X} \prod_i \left[ \frac{\alpha_i \left(M_{j+1}\right)}{\alpha_i \left(M_{j}\right)} \right]^{\frac{\gamma_i}{b_i}} ,
\end{align}
where, as usual, $\alpha_i = g_i^2/4\pi $ and $\gamma_i$'s are the relevant anomalous dimensions, values of which are  shown in Table~\ref{table_anomalous}. Whereas the  $\beta$-coefficients $b_i$ are computed in the next subsection.

\begin{table}[h!]
	\renewcommand*{\arraystretch}{1.2}
	\begin{center}
		\begin{tabular}{|c|c|c|}
			\hline
			\multirow{2}{*}{Gauge group} & \multicolumn{2}{c|}{Anomalous dimensions}\\
			\cline{2-3}
			& $\mathcal{O}^{d=6}_L\left(e^C , d\right)$ & $\mathcal{O}^{d=6}_R\left(e , d^C \right)$  \\ 
			\hline
			$G_{321}$& $\lbrace 2, \frac{9}{4}, \frac{23}{20} \rbrace$ & $\lbrace 2, \frac{9}{4}, \frac{11}{20}  \rbrace$  \\
			\hline
			$G_{3221}$& $\lbrace 2, \frac{9}{4}, \frac{9}{4}, \frac{1}{4} \rbrace$ & $\lbrace 2, \frac{9}{4}, \frac{9}{4}, \frac{1}{4} \rbrace$ \\
			\hline
			$G_{422}$& $\lbrace \frac{15}{4}, \frac{9}{4}, \frac{9}{4} \rbrace$ &  $\lbrace \frac{15}{4}, \frac{9}{4}, \frac{9}{4}  \rbrace$ \\

			\hline
		\end{tabular}
		\caption{Relevant anomalous dimensions for the considered breaking chains.}
		\label{table_anomalous}
	\end{center}
	
\end{table}

\subsection{Unification}
The two-loop renormalization group equations (RGE) for the gauge couplings can be written as,
\begin{align}
&\dfrac{d \alpha^{-1}_i(\hat\mu)}{d \ln \hat\mu}=-\dfrac{b_i}{2 \pi}-\sum\limits_{j}\dfrac{b_{ij}}{8 \pi^2 \alpha^{-1}_j(\hat\mu)}, 
\end{align}
where $i,j$ indices refer to different daughter groups of the parent group at the energy scale $\hat\mu$ and we have defined,
\begin{align}
&\alpha^{-1}_i=\dfrac{4 \pi}{g_i^2}.
\end{align}
Using the generalized formula for the beta-function, $\beta = \hat\mu d g/d\hat\mu$, one finds~\cite{Machacek:1983tz},
\begin{align}
b_i&=-\dfrac{11}{3}C_2 (G_i)+\dfrac{4}{3}\kappa S_2(F_i)+\dfrac{1}{6}\eta S_2(S_i),\label{bi}\\
b_{ij}&=-\dfrac{34}{3} \left[ C_2(G_i)\right] ^2 \delta_{ij}+\kappa\left[4 C_2(F_j) +\dfrac{20}{3}\delta_{ij} C_2(G_i)\right]S_2(F_i) \nonumber\\ 
&+\eta \left[2C_2(S_j)+\dfrac{1}{3} \delta_{ij}  C_2(G_i)\right]S_2(S_i).\label{bij} \end{align}
Here $S_2$ and $C_2$ are the Dynkin indices of the representations with the appropriate multiplicity factors and the quadratic Casimir of a given representation. For Dirac and Weyl fermions $\kappa=1$ and $\frac{1}{2}$, respectively, and $\eta=1,2$ for real and complex scalar fields. $G$, $F$ and $S$ stand for gauge multiplets, fermions, and scalars.

The decomposition of the $SO(10)$ representations under the relevant sub-groups are  presented in Tables~\ref{tab:10decomposition}, \ref{tab:45decomposition}, \ref{tab:54decomposition}, \ref{tab:120decomposition}, and \ref{tab:126decomposition}. In these tables,  multiplets highlighted in red, blue, and green contribute to the RGE running of the gauge coupling in the energy scale in between $SO(10)$ and $G_{422}$, $G_{422}$ and $G_{3221}$, and $G_{3221}$ and $G_{321}$, respectively. On the other hand, multiplets highlighted in gray represent the Goldstone bosons. Utilizing these information from Tables~\ref{tab:10decomposition}-\ref{tab:126decomposition} and using Eqs.~\eqref{bi}-\eqref{bij}, in the following, we compute the relevant beta coefficients required for gauge coupling unification.

The beta coefficients of the SM sector are well-known, which are:
\begin{align}
&\bigg\{b^\textrm{SM}_{1Y}, b^\textrm{SM}_{2L}, b^\textrm{SM}_{3C}  \bigg\}=\left\{\frac{41}{10},-\frac{19}{6},-7\right\}, 
\\
&b^\textrm{SM}_{ij}=\left(
\begin{array}{ccc}
 \frac{199}{50} & \frac{27}{10} & \frac{44}{5} \\
 \frac{9}{10} & \frac{35}{6} & 12 \\
 \frac{11}{10} & \frac{9}{2} & -26 \\
\end{array}
\right).
\end{align}
For our model, the beta coefficients relevant to the energy range between left-right symmetry and Pati-Salam symmetry are:
\begin{align}
&\bigg\{b^\textrm{LR}_{2L}, b^\textrm{LR}_{2R}, b^\textrm{LR}_{3C}, b^\textrm{LR}_{BL}  \bigg\}=\left\{-\frac{17}{6},-\frac{13}{6},-7,\frac{11}{2}\right\},  
\\
&b^\textrm{LR}_{ij}=
\left(
\begin{array}{cccc}
 \frac{61}{6} & \frac{9}{2} & 12 & \frac{3}{2} \\
 \frac{9}{2} & \frac{173}{6} & 12 & \frac{27}{2} \\
 \frac{9}{2} & \frac{9}{2} & -26 & \frac{1}{2} \\
 \frac{9}{2} & \frac{81}{2} & 4 & \frac{61}{2} \\
\end{array}
\right),
\end{align}
and in the energy range between  Pati-Salam and GUT symmetries are:
\begin{align}
&\bigg\{b^\textrm{PS}_{2L}, b^\textrm{PS}_{2R}, b^\textrm{PS}_{4C}  \bigg\}=\left\{\frac{17}{2},\frac{17}{2},\frac{4}{3}\right\},  
\\
&b^\textrm{PS}_{ij}=\left(
\begin{array}{ccc}
 \frac{515}{2} & \frac{93}{2} & \frac{1245}{2} \\
 \frac{93}{2} & \frac{515}{2} & \frac{1245}{2} \\
 \frac{249}{2} & \frac{249}{2} & \frac{3775}{6} \\
\end{array}
\right).
\end{align}

\begin{figure}[t!]
\centering
\includegraphics[width=1\textwidth]{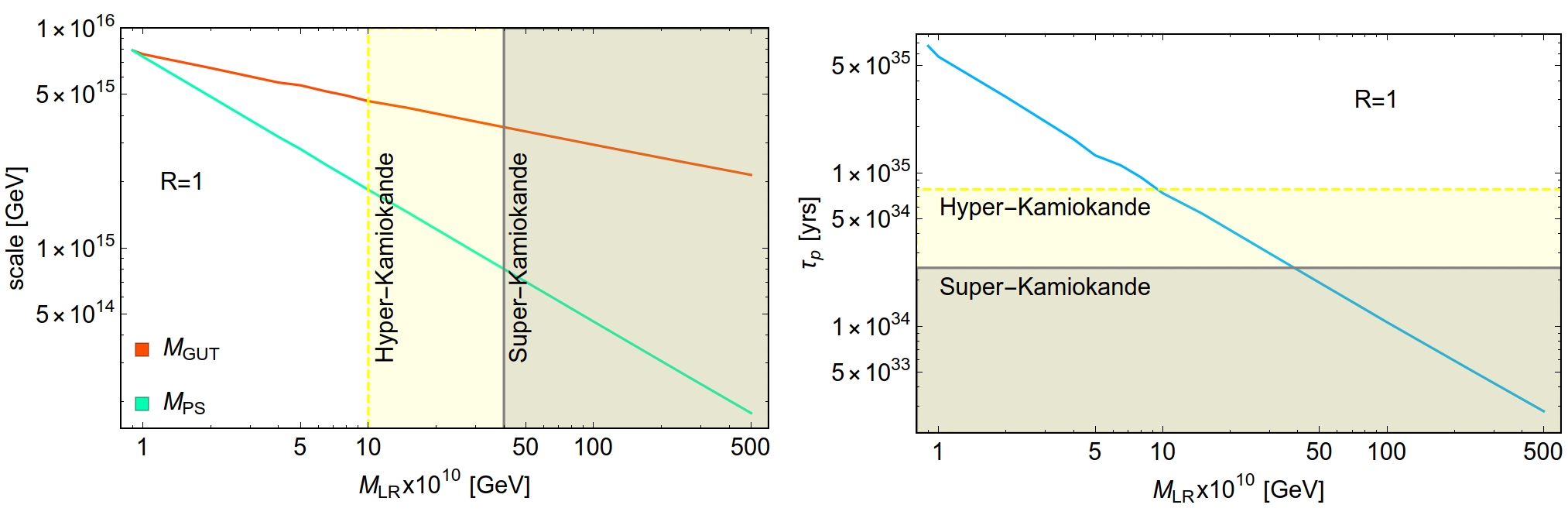}
\caption{Plot without incorporating threshold correction, i.e., $R=1$. Current Super-K bound requires $M_{II}< 4\times 10^{11}$ GeV, and future Hyper-K bound would push this scale to below $M_{II}<10^{11}$ GeV.  } \label{fig:R1}
\end{figure}

At a symmetry breaking scale, $\hat\mu$, transitioning from the parent group $G_p$ to the daughter gauge group $G_d$ happens. This requires a proper matching condition between the gauge couplings $\alpha_d^{-1}$ of the daughter gauge group and the gauge couplings $\alpha_p^{-1}$ of the parent group, which reads
\begin{align}
\alpha_d^{-1}(\hat\mu)-\frac{C_2(G_d)}{12\pi}=\left(\alpha_p^{-1}(\hat\mu) -\frac{C_2(G_p)}{12\pi}\right) - \frac{\Lambda_d(\hat\mu)}{12\pi} , 
\end{align}
where one-loop threshold corrections after integrating out the superheavy fields are given by~\cite{Weinberg:1980wa,Hall:1980kf,Bertolini:2009qj,Bertolini:2013vta},
\begin{align}
\Lambda_d(\hat\mu)=-21\; {\mathrm{Tr}} (t_{dV}^2 \ln\frac{m_V}{\hat\mu})+2\;\eta\; {\mathrm{Tr}} (t_{dS}^2 \ln\frac{m_S}{\hat\mu}) + 8\;\kappa\; {\mathrm{Tr}} (t_{dF}^2 \ln\frac{m_F}{\hat\mu}).
\end{align}
Here, $t_{dV}$, $t_{dS}$, and $t_{dF}$ are the generators for the representations under $G_d$ of the superheavy vector, scalar, and fermion fields, respectively, $m_V$, $m_S$, and $m_F$ are their respective masses.

First, we consider a scenario without threshold corrections, meaning we set $R=m_S/\hat\mu=1$; the result is presented in Fig.~\ref{fig:R1}. As can be seen from this figure, the current Super-K~\cite{Super-Kamiokande:2020wjk} bound demands the seesaw scale to be smaller than $M_{II}< 4\times 10^{11}$ GeV. On the other hand, future Hyper-K~\cite{Hyper-Kamiokande:2018ofw} bound would push this scale to below $M_{II}<10^{11}$ GeV, leaving a small window for which gauge coupling unification can be obtained.  However, the seesaw scale of this magnitude is inconsistent with the findings of the PTAs and the fermion mass fit described above. This shows the need for considering  the threshold corrections.

\begin{figure}[t!]
\centering
\includegraphics[width=1\textwidth]{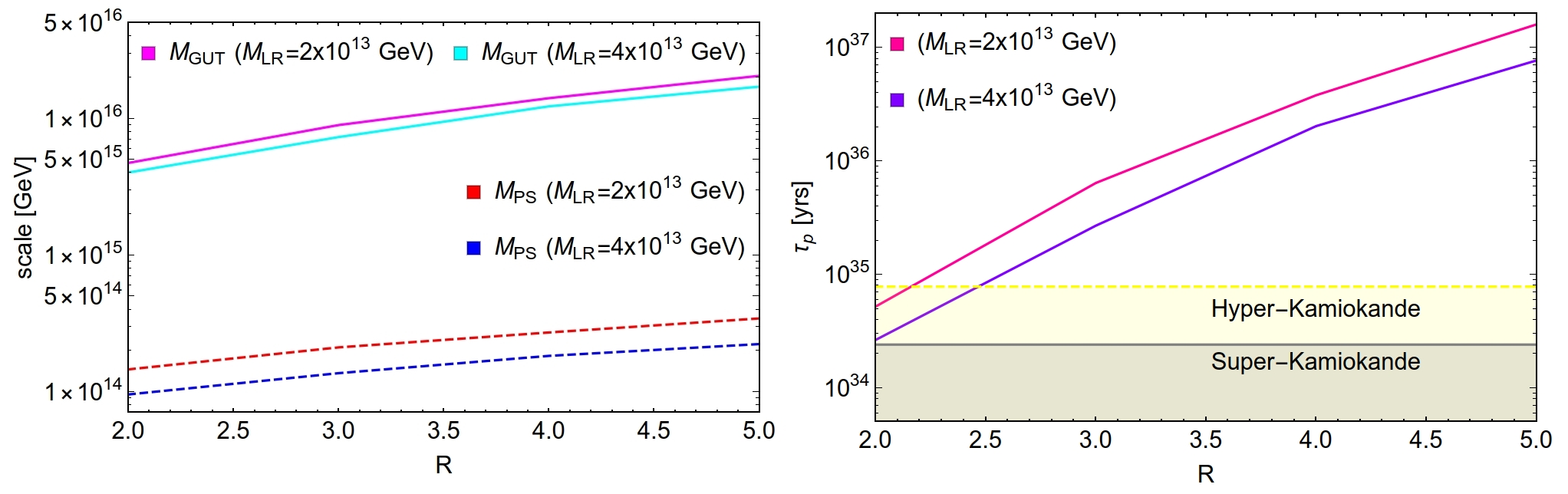}
\caption{ Plot with threshold corrections for $R=[2-5]$.  As can be seen from this plot, a threshold correction with $R=2$ is enough to achieve high scale gauge coupling unification which is also consistent with the current lower bound on proton decay. Hence no additional fine-turning is required in this model.   } \label{fig:R5}
\end{figure}

When threshold corrections are turned on, we follow the common approach in which the relevant superheavy gauge bosons at each symmetry breaking scale have the masses identified with the corresponding breaking scale. On the other hand, the scalars are assumed to be non-degenerate; hence, only scalars contribute to the threshold correction.  Masses of the scalars are varied within the range $[1/R-R]$, with $R=2-5$.  In this procedure, all the superheavy degrees of freedom have the masses around the breaking scale within the span $[1/R-R]$. We perform a $\chi^2$-analysis with a fixed seesaw scale (corresponding to $M_\mathrm{LR}=2\times 10^{13}$ GeV and $4\times 10^{13}$ GeV), where the GUT scale is maximized by varying the associated threshold corrections within a given range.  The result is depicted in Fig.~\ref{fig:R5}. As can be seen from this figure, allowing a small threshold correction is enough for the compatibility with gravitational wave observations, fermion masses and mixings, as well as unification and proton decay constraints.

\section{Conclusions}\label{sec:conclusions}
Grand unified theories   are attractive candidates for physics beyond the Standard Model.  
In this work, we have considered a minimal realistic grand unified theory based on $SO(10)$ gauge group. The Yukawa sector of this theory consists of a real $10_H$, a real $120_H$, and a complex $126_H$ dimensional representations, which is the most minimal option without introducing any additional symmetries. The symmetry breaking sector that employs a real $54_H$ and a real $45_H$ dimensional representations is also taken to be the most economical that can give rise to stable cosmic strings when the last step of the symmetry breaking takes place. Intriguingly,  a series of pulsar-timing arrays  might have already observed GWs in the nHz regime, hinting towards forming a cosmic string network in the very early universe, which we identified with the seesaw scale generating tiny masses for the active neutrinos. To be consistent with these observations, we find that the seesaw scale is restricted to be within a narrow range $v_R\in(3.4,4.1)\times 10^{13}$ GeV. More data, however, is required
to confirm if these signals are from a stochastic gravitational wave background.

Since within this framework, the charged fermion masses are inherently interconnected with the neutrino masses,  low energy data of fermion masses and mixing angles allow this seesaw scale in the range $v_R\in(10^{13}, 10^{15})$ GeV. Contrastingly, if no threshold correction from the scalar spectrum is considered, constraints from gauge coupling unification and proton decay demand a much smaller  seesaw scale $v_R\in(10^{10}, 10^{11})$ GeV. This latter range for the seesaw scale is in direct conflict with observations of pulsar-timing arrays as well as from low energy data. Our detailed study shows that if the scalars are not assumed to be degenerate with the corresponding symmetry breaking scales, then a small amount of threshold correction is sufficient for compatibility with gravitational wave observations, fermion mass fit, gauge coupling unification, and proton decay bounds.  The proposed minimal model is highly predictive and will be fully tested in a series of experiments in the near future.

\section*{Acknowledgments}
S.S\; would like to thank Valerie Domcke for useful discussion.

\begin{appendices}
\renewcommand\thesection{\arabic{section}}
\renewcommand\thesubsection{\thesection.\arabic{subsection}}
\renewcommand\thefigure{\arabic{figure}}
\renewcommand\thetable{\arabic{table}}

\section{Best fit parameters without restricting the seesaw scale}\label{A}

The best fit parameters for the NO solution is as follows: 
\begin{align}
&r_1 = 4.99994\times 10^{-3}\, +6.64062\times 10^{-5} i, \label{fit1:NO} \\
&r_2= -1.33796 -0.257534 i, \\
&\phi=-0.0504299,  \\
&c_R= 2.72334\times 10^{12}, 
\end{align}

\begin{equation}
S= \left(
\begin{array}{ccc}
7.83123\times 10^{-8} & 0 & 0 \\
 0 & 0.237256 & 0 \\
 0 & 0 & 83.9454 \\
\end{array}
\right) \;\rm{GeV},
\end{equation}

\begin{equation}
D=\scalemath{0.9}
{ \left(
\begin{array}{ccc}
 0.00051669\, +0.000140812 i & 0.00470802\, +0.00142105 i & 0.0293596\, +0.00308667 i \\
 0.00470802\, +0.00142105 i & 0.0299984\, -0.0299305 i & 0.457075\, -0.132563 i \\
 0.0293596\, +0.00308667 i & 0.457075\, -0.132563 i & 0.0118382\, -0.456846 i \\
\end{array}
\right)}\;\rm{GeV},
\end{equation}

\begin{equation}
A=\scalemath{0.9}
{
\left(
\begin{array}{ccc}
 0 & 0.00363066\, +0.00172004 i & 0.0220297\, +0.00302919 i \\
 -0.00363066-0.00172004 i & 0 & 0.42873\, -0.0919318 i \\
 -0.0220297-0.00302919 i & -0.42873+0.0919318 i & 0 \\
\end{array}
\right)} \;\rm{GeV}. \label{fit2:NO}
\end{equation}

For the IO case, the best fit parameters are given below,
\begin{align}
&r_1 = 5.26833\times 10^{-3}\, -3.51051\times 10^{-4} i, \label{fit1:IO} \\
&r_2= -1.42403 - 0.253046 i, \\
&\phi=0.0569906,  \\
&c_R= 2.82679\times 10^{12}, 
\end{align}

\begin{equation}
S= \left(
\begin{array}{ccc}
 4.62836\times 10^{-9} & 0 & 0 \\
 0 & 0.227095 & 0 \\
 0 & 0 & 83.8975 \\
\end{array}
\right) \;\rm{GeV},
\end{equation}

\begin{equation}
D=\scalemath{0.88}
{ \left(
\begin{array}{ccc}
 0.000544454\, +0.0000187733 i & 0.00466318\, +6.02921\times 10^{-6} i & 0.0232066\, -0.0017959
   i \\
 0.00466318\, +6.02921\times 10^{-6} i & 0.0417626\, -0.0141477 i & 0.497413\, -0.00289039 i \\
 0.0232066\, -0.0017959 i & 0.497413\, -0.00289039 i & 0.0700564\, -0.185603 i \\
\end{array}
\right)}\;\rm{GeV},
\end{equation}

\begin{equation}
A=\scalemath{0.9}
{
\left(
\begin{array}{ccc}
 0 & 0.00322916\, +0.000429876 i & 0.0162086\, -0.000403031 i \\
 -0.00322916-0.000429876 i & 0 & 0.460026\, -0.0329099 i \\
 -0.0162086+0.000403031 i & -0.460026+0.0329099 i & 0 \\
\end{array}
\right)}\;\rm{GeV}. \label{fit2:IO}
\end{equation}

\section{Best fit parameters for fixed seesaw scale}\label{B}
As described in the main text, in achieving this fit, we fixed $M_3=4\times 10^{13}$ GeV. The best fit parameters in this scenario are given by, 
\begin{align}
&r_1 = 4.79222\times 10^{-3}\, +5.85329\times 10^{-5} i, \label{fit1:fix} \\
&r_2= -0.88177 - 1.51667 i, \\
&\phi=0.682943,  \\
&c_R= 4.7\times 10^{11},
\end{align}

\begin{equation}
S= \left(
\begin{array}{ccc}
 9.97395\times 10^{-7} & 0 & 0 \\
 0 & 0.28853 & 0 \\
 0 & 0 & 85.6185 \\
\end{array}
\right) \;\rm{GeV},
\end{equation}

\begin{equation}
D= \scalemath{0.88}
{\left(
\begin{array}{ccc}
 -0.000178778+0.000261172 i & 0.00621503\, +0.000737064 i & 0.0124298\, -0.000291788 i
   \\
 0.00621503\, +0.000737064 i & -0.0135252+0.0153914 i & 0.337599\, +0.0444584 i \\
 0.0124298\, -0.000291788 i & 0.337599\, +0.0444584 i & 0.0394304\, -0.832908 i \\
\end{array}
\right)} \;\rm{GeV},
\end{equation}

\begin{equation}
A=\scalemath{0.88}
{
\left(
\begin{array}{ccc}
 0 & 0.00297024\, +0.00195566 i & 0.0115613\, -0.0108631 i \\
 -0.00297024-0.00195566 i & 0 & 0.235703\, -0.157565 i \\
 -0.0115613+0.0108631 i & -0.235703+0.157565 i & 0 \\
\end{array}
\right)} \;\rm{GeV}. \label{fit2:fix}
\end{equation}

\section{Decomposition of representations}\label{C}
The decomposition of the $SO(10)$ representations under the relevant sub-groups are  presented in Tables~\ref{tab:10decomposition}, \ref{tab:45decomposition}, \ref{tab:54decomposition}, \ref{tab:120decomposition}, and \ref{tab:126decomposition}.

\FloatBarrier
\begin{table}[th!]
\centering
\footnotesize
\resizebox{1\textwidth}{!}{
\begin{tabular}{|c|c|c|}
\hline

\textbf{$SU(4)_C\times SU(2)_L\times SU(2)_R$}  & \textbf{$SU(3)_C\times SU(2)_L\times SU(2)_R\times U(1)_{B-L}$} & \textbf{$SU(3)_C\times SU(2)_L\times U(1)_Y$}  \\ 

\hline \hline

$\mathbf{(6,1,1)}$ & $\mathbf{(3,1,1,-2/3)}$ & $\mathbf{(3,1,-1/3)}$\\  
 &$\mathbf{(\overline 3,1,1,2/3)}$ & $\mathbf{(\overline 3,1,1/3)}$\\ \hline \hline

 \tikz[baseline]{\node[fill=red!20,anchor=base]{$\mathbf{(1,2,2)}$}}&\tikz[baseline]{\node[fill=blue!20,anchor=base]{$\mathbf{(1,2,2,0)}$}} & $\mathbf{(1,2,1/2)}$\\
 & & $\mathbf{(1,2,-1/2)}$ \\  \hline

\end{tabular}
}
\caption{ Decomposition of $\mathbf{10}$-Higgs, which is a real field.  Multiplets highlighted in red, blue, and green contribute to the RGE running of the gauge coupling in the energy scale in between $SO(10)$ and $G_{422}$, $G_{422}$ and $G_{3221}$, and $G_{3221}$ and $G_{321}$, respectively. Multiplets highlighted in gray represent the Goldstone bosons.  }\label{tab:10decomposition}
\end{table}

\FloatBarrier
\begin{table}[th!]
\centering
\footnotesize
\resizebox{1\textwidth}{!}{
\begin{tabular}{|c|c|c|}
\hline

\textbf{$SU(4)_C\times SU(2)_L\times SU(2)_R$}  & \textbf{$SU(3)_C\times SU(2)_L\times SU(2)_R\times U(1)_{B-L}$} & \textbf{$SU(3)_C\times SU(2)_L\times U(1)_Y$}  \\ 

\hline \hline

 $\mathbf{(1,1,3)}$ & $\mathbf{(1,1,3,0)}$ & $\mathbf{(1,1,1)}$\\  
 & & $\mathbf{(1,1,0)}$\\
 & & $\mathbf{(1,1,-1)}$\\ \hline \hline

 $\mathbf{(1,3,1)}$&$\mathbf{(1,3,1,0)}$ & $\mathbf{(1,3,0)}$\\ \hline \hline

 $\mathbf{(6,2,2)}$ & $\mathbf{(3,2,2,-2/3)}$ & $\mathbf{(3,2,1/6)}$\\  
 & & $\mathbf{(3,2,-5/6)}$\\
  & $\mathbf{(\overline 3,2,2,2/3)}$ & $\mathbf{(\overline 3,2,5/6)}$\\  
 & & $\mathbf{(\overline 3,2,-1/6)}$\\ \hline \hline

 \tikz[baseline]{\node[fill=red!20,anchor=base]{$\mathbf{(15,1,1)}$}} & $\mathbf{(1,1,1,0)}$ & $\mathbf{(1,1,0)}$\\  
 &\tikz[baseline]{\node[fill=gray!20,anchor=base]{$\mathbf{(3,1,1,4/3)}$}} & $\mathbf{(3,1,2/3)}$\\
  &\tikz[baseline]{\node[fill=gray!20,anchor=base]{$\mathbf{(\overline 3,1,1,-4/3)}$}} & $\mathbf{(\overline 3,1,-2/3)}$\\  
 &$\mathbf{(8,1,1,0)}$ & $\mathbf{(8,1,0)}$\\ \hline 

\end{tabular}
}
\caption{ Decomposition of $\mathbf{45}$-Higgs, which is a real field. Multiplets highlighted in red, blue, and green contribute to the RGE running of the gauge coupling in the energy scale in between $SO(10)$ and $G_{422}$, $G_{422}$ and $G_{3221}$, and $G_{3221}$ and $G_{321}$, respectively. Multiplets highlighted in gray represent the Goldstone bosons.  }\label{tab:45decomposition}
\end{table}

\FloatBarrier
\begin{table}[th!]
\centering
\footnotesize
\resizebox{1\textwidth}{!}{
\begin{tabular}{|c|c|c|}
\hline

\textbf{$SU(4)_C\times SU(2)_L\times SU(2)_R$}  & \textbf{$SU(3)_C\times SU(2)_L\times SU(2)_R\times U(1)_{B-L}$} & \textbf{$SU(3)_C\times SU(2)_L\times U(1)_Y$}  \\ 

\hline \hline

 $\mathbf{(1,1,1)}$ & $\mathbf{(1,1,1,0)}$ & $\mathbf{(1,1,0)}$\\   \hline \hline

 $\mathbf{(1,3,3)}$ & $\mathbf{(1,3,3,0)}$ & $\mathbf{(1,3,1)}$\\  
 & & $\mathbf{(1,3,0)}$\\
 & & $\mathbf{(1,3,-1)}$\\ \hline \hline

 $\mathbf{(20^{\prime},1,1)}$ & $\mathbf{(\overline 6,1,1,4/3)}$ & $\mathbf{(\overline 6,1,2/3)}$\\  
 & $\mathbf{(6,1,1,-4/3)}$ & $\mathbf{(6,1,-2/3)}$\\
 & $\mathbf{(8,1,1,0)}$ & $\mathbf{(8,1,0)}$\\ \hline \hline

\tikz[baseline]{\node[fill=gray!20,anchor=base]{$\mathbf{(6,2,2)}$}} & $\mathbf{(3,2,2,-2/3)}$ & $\mathbf{(3,2,1/6)}$\\  
 & & $\mathbf{(3,2,-5/6)}$\\
  & $\mathbf{(\overline 3,2,2,2/3)}$ & $\mathbf{(\overline 3,2,5/6)}$\\  
 & & $\mathbf{(3,2,-1/6)}$\\  \hline

\end{tabular}
}
\caption{ Decomposition of $\mathbf{54}$-Higgs, which is a real field. Multiplets highlighted in red, blue, and green contribute to the RGE running of the gauge coupling in the energy scale in between $SO(10)$ and $G_{422}$, $G_{422}$ and $G_{3221}$, and $G_{3221}$ and $G_{321}$, respectively. Multiplets highlighted in gray represent the Goldstone bosons.  }\label{tab:54decomposition}
\end{table}

\FloatBarrier
\begin{table}[th!]
\centering
\footnotesize
\resizebox{1\textwidth}{!}{
\begin{tabular}{|c|c|c|}
\hline

\textbf{$SU(4)_C\times SU(2)_L\times SU(2)_R$}  & \textbf{$SU(3)_C\times SU(2)_L\times SU(2)_R\times U(1)_{B-L}$} & \textbf{$SU(3)_C\times SU(2)_L\times U(1)_Y$}  \\ 

\hline \hline

 $\mathbf{(1,2,2)}$ & $\mathbf{(1,2,2,0)}$ & $\mathbf{(1,2,1/2)}$\\
   &  & $\mathbf{(1,2,-1/2)}$\\\hline \hline

 $\mathbf{(10,1,1)}$ & $\mathbf{(1,1,1,-2)}$ & $\mathbf{(1,1,-1)}$\\  
 & $\mathbf{(3,1,1,-2/3)}$ & $\mathbf{(3,1,-1/3)}$\\
 & $\mathbf{(6,1,1,2/3)}$ & $\mathbf{(6,1,1/3)}$\\ \hline \hline

 $\mathbf{(\overline 10,1,1)}$ & $\mathbf{(1,1,1,2)}$ & $\mathbf{(1,1,1)}$\\  
 & $\mathbf{(\overline 3,1,1,2/3)}$ & $\mathbf{(\overline 3,1,1/3)}$\\
 & $\mathbf{(\overline 6,1,1,-2/3)}$ & $\mathbf{(\overline 6,1,-1/3)}$\\ \hline \hline

 $\mathbf{(6,3,1)}$ & $\mathbf{(3,3,1,-2/3)}$ & $\mathbf{(3,3,-1/3)}$\\  
 & $\mathbf{(\overline 3,3,1,2/3)}$ & $\mathbf{(\overline 3,3,1/3)}$\\ \hline \hline

 $\mathbf{(6,1,3)}$ & $\mathbf{(3,1,3,-2/3)}$ & $\mathbf{(3,1,2/3)}$\\ 
 &  & $\mathbf{(3,1,-1/3)}$\\ 
  &  & $\mathbf{(3,1,-4/3)}$\\ 
 & $\mathbf{(\overline 3,1,3,2/3)}$ & $\mathbf{(\overline 3,1,4/3)}$\\ 
 &  & $\mathbf{(\overline 3,1,1/3)}$\\ 
 &  & $\mathbf{(\overline 3,1,-2/3)}$\\  \hline \hline

 $\mathbf{(15,2,2)}$ & $\mathbf{(1,2,2,0)}$ & $\mathbf{(1,2,1/2)}$\\ 
&  & $\mathbf{(1,2,-1/2)}$\\ 
 & $\mathbf{(3,2,2,4/3)}$  & $\mathbf{(3,2,7/6)}$\\ 
 &  & $\mathbf{(\overline 3,2,1/6)}$\\ 
 & $\mathbf{(\overline 3,2,2,-4/3)}$  & $\mathbf{(\overline 3,2,-7/6)}$\\ 
 &  & $\mathbf{(3,2,-1/6)}$\\
  & $\mathbf{(8,2,2,0)}$ & $\mathbf{(8,2,1/2)}$\\ 
 &  & $\mathbf{(8,2,-1/2)}$\\ \hline \hline

\end{tabular}
}
\caption{ Decomposition of $\mathbf{120}$-Higgs, which is a real field. For the simplicity of the analysis, we assume the entire $120_H$ lives near to the GUT scale and does not contribute to the RGEs. Putting the bi-doublet at the intermediate scale, however, can be straightforwardly incorporated.  }\label{tab:120decomposition}
\end{table}

\FloatBarrier
\begin{table}[th!]
\centering
\footnotesize
\resizebox{1\textwidth}{!}{
\begin{tabular}{|c|c|c|}
\hline

\textbf{$SU(4)_C\times SU(2)_L\times SU(2)_R$}  & \textbf{$SU(3)_C\times SU(2)_L\times SU(2)_R\times U(1)_{B-L}$} & \textbf{$SU(3)_C\times SU(2)_L\times U(1)_Y$}  \\ 

\hline \hline

$\mathbf{(6,1,1)}$ & $\mathbf{(3,1,1,-2/3)}$ & $\mathbf{(3,1,-1/3)}$\\  
 &$\mathbf{(\overline 3,1,1,2/3)}$ & $\mathbf{(\overline 3,1,1/3)}$\\ \hline \hline

 \tikz[baseline]{\node[fill=red!20,anchor=base]{$\mathbf{(\overline{10},3,1)}$}} & $\mathbf{(1,3,1,2)}$ & $\mathbf{(1,3,1)}$\\  
 &$\mathbf{(\overline 3,3,1,2/3)}$ & $\mathbf{(\overline 3,3,1/3)}$\\ 
 &$\mathbf{(\overline 6,3,1,-2/3)}$ & $\mathbf{(\overline 6,3,-1/3)}$\\ \hline \hline

 \tikz[baseline]{\node[fill=red!20,anchor=base]{$\mathbf{(10,1,3)}$}} & \tikz[baseline]{\node[fill=blue!20,anchor=base]{$\mathbf{(1,1,3,-2)}$}} & \tikz[baseline]{\node[fill=gray!20,anchor=base]{$\mathbf{(1,1,0)}$}}\\ 
 &  & \tikz[baseline]{\node[fill=gray!20,anchor=base]{$\mathbf{(1,1,-1)}$}}\\ 
&  & $\mathbf{(1,1,-2)}$\\ 
  & $\mathbf{(3,1,3,-2/3)}$ & $\mathbf{(3,1,2/3)}$\\ 
 &  & $\mathbf{(3,1,-1/3)}$\\ 
 &  & $\mathbf{(3,1,-4/3)}$\\ 
 & $\mathbf{(6,1,3,2/3)}$ & $\mathbf{(6,1,4/3)}$\\ 
 &  & $\mathbf{(6,1,1/3)}$\\ 
 &  & $\mathbf{(6,1,-2/3)}$\\   \hline \hline

 \tikz[baseline]{\node[fill=red!20,anchor=base]{$\mathbf{(15,2,2)}$}} & \tikz[baseline]{\node[fill=blue!20,anchor=base]{$\mathbf{(1,2,2,0)}$}} & \tikz[baseline]{\node[fill=green!20,anchor=base]{$\mathbf{(1,2,1/2)}$}}\\ 
 &  & \tikz[baseline]{\node[fill=green!20,anchor=base]{$\mathbf{(1,2,-1/2)}$}}\\ 
 & $\mathbf{(3,2,2,4/3)}$  & $\mathbf{(3,2,7/6)}$\\ 
&  & $\mathbf{(\overline 3,2,1/6)}$\\ 
 & $\mathbf{(\overline 3,2,2,-4/3)}$  & $\mathbf{(\overline 3,2,-7/6)}$\\ 
 &  & $\mathbf{(3,2,-1/6)}$\\
  & $\mathbf{(8,2,2,0)}$ & $\mathbf{(8,2,1/2)}$\\ 
&  & $\mathbf{(8,2,-1/2)}$\\ \hline \hline

\end{tabular}
}
\caption{ Decomposition of $\mathbf{\overline{126}}$-Higgs, which is a complex field. Multiplets highlighted in red, blue, and green contribute to the RGE running of the gauge coupling in the energy scale in between $SO(10)$ and $G_{422}$, $G_{422}$ and $G_{3221}$, and $G_{3221}$ and $G_{321}$, respectively. Multiplets highlighted in gray represent the Goldstone bosons.  }\label{tab:126decomposition}
\end{table}

\end{appendices}

\clearpage
\bibliographystyle{style}
\bibliography{reference}
\end{document}